\documentclass[aps,twocolumn,prx,floatfix,showpacs,lengthcheck,citeautoscript,superscriptaddress,longbibliography]{revtex4-2}
\usepackage{amsmath}
\usepackage{amssymb}
\usepackage{graphicx}
\usepackage{bm}
\usepackage{color,soul}
\usepackage{braket}
\usepackage[dvipsnames,svgnames,table]{xcolor}
\usepackage{times}
\usepackage{MnSymbol}
\usepackage{float}
\usepackage{bbold}
\usepackage{babel}
\usepackage{hyperref}
\usepackage{orcidlink}
\usepackage[capitalise]{cleveref}
\definecolor{Nathanpink}{rgb}{0.94,0.317,0.9607}

\definecolor{titles}{rgb}{0.,0.24,0.51}

\hypersetup{
pdfstartview={FitH},
colorlinks=true,    
linkcolor=NavyBlue, 
citecolor=NavyBlue,   
filecolor=NavyBlue, 
urlcolor=NavyBlue   
}

\newcommand{\smeq}{\! = \!}

\newcommand{\ve}{\varepsilon}

\newcommand{\bea}{\begin{eqnarray}}
\newcommand{\eea}{\end{eqnarray}}
\newcommand{\be}{\begin{equation}}
\newcommand{\ee}{\end{equation}}

\newcommand{\la}{\langle}
\newcommand{\ra}{\rangle}

\let\Im\relax
\DeclareMathOperator{\Im}{Im}

\begin{document}
 
\title{Luttinger's Theorem Violation and Green's Function Topological Invariants\\
in a Fractional Chern Insulator}

\author{Anton A. Markov\textsuperscript{\P}}
\email{anton.markov@ulb.be}
\affiliation{International Solvay Institutes, 1050 Brussels, Belgium}
\affiliation{Center for Nonlinear Phenomena and Complex Systems, Université Libre de Bruxelles (U.L.B.), B-1050 Brussels, Belgium}
\affiliation{Russian Quantum Center, Moscow 121205, Russia}
\thanks{These authors contributed equally to this work.}

\author{Andrey M. Nikishin\textsuperscript{\P}}
\email{a.nikishin@rqc.ru}
\affiliation{Russian Quantum Center, Moscow 121205, Russia}
\affiliation{Moscow Institute of Physics and Technology, Institutsky lane 9, Dolgoprudny, Moscow region,
141700}

\author{Nigel R. Cooper}
\affiliation{T.C.M. Group, Cavendish Laboratory, University of Cambridge, J.J. Thomson Avenue, Cambridge CB3 0US, United Kingdom}

\author{Nathan Goldman}
\email{nathan.goldman@lkb.ens.fr}
\affiliation{International Solvay Institutes, 1050 Brussels, Belgium}
\affiliation{Center for Nonlinear Phenomena and Complex Systems, Université Libre de Bruxelles (U.L.B.), B-1050 Brussels, Belgium}
\affiliation{Laboratoire Kastler Brossel, Coll\`ege de France, CNRS, ENS-Universit\'e PSL, Sorbonne Universit\'e, 11 Place Marcelin Berthelot, 75005 Paris, France}

\author{Lucila Peralta Gavensky}
\email{lucila.peralta.gavensky@ulb.be}
\affiliation{International Solvay Institutes, 1050 Brussels, Belgium}
\affiliation{Center for Nonlinear Phenomena and Complex Systems, Université Libre de Bruxelles (U.L.B.), B-1050 Brussels, Belgium}

\begin{abstract}
    
Luttinger's theorem constrains the particle density of interacting fermions through global properties of the single-particle Green's function, and its violation signals a breakdown of the identification between the quantized Hall response and the Green-function-based Ishikawa-Matsuyama invariant. This phenomenon becomes especially compelling in strongly correlated topological phases, such as fractional Chern insulators, where fractionalized quasiparticles lack an adiabatic connection to electrons, raising the question of how Green's-function-based topological invariants manifest in such phases.
Using exact diagonalization of the fermionic Harper-Hofstadter-Hubbard model, we compute bulk single-particle Green's functions deep inside a fractional Chern insulating phase and directly evaluate the Luttinger count, its possible correction (the Luttinger integral), and the Ishikawa-Matsuyama invariant $N_3[\mathrm{G}]$. We demonstrate a clear violation of Luttinger's theorem and show that the fractional nature of the many-body Chern number is encoded in the St\v{r}eda response of the Luttinger integral, while the integer invariant 
$N_3[\mathrm{G}]$
 arises from the St\v{r}eda response of 
the Luttinger count. We also analytically prove that
$N_3[\mathrm{G}]$ is fully determined by the Luttinger count together with the Chern number of the occupied Bloch band, upon neglecting Bloch-band mixing. Finally, we propose an experimental protocol to extract all Green-function-based topological invariants from local density-of-states measurements, experimentally accessible in fractional quantum Hall systems.
    
\end{abstract}

\date{\today}

\bibliographystyle{unsrt}
\maketitle
\section{Introduction}

The success of many-body perturbation theory relies on the assumption of adiabatic continuity between electrons and low-energy quasiparticle excitations \cite{lifshitz2013statistical, luttinger1960ground}. In such cases, the single-particle Green's function, which describes electron propagation through a many-body system, encodes detailed information about these low-energy degrees of freedom. The situation is markedly different in Fractional Quantum Hall (FQH) \cite{tsui1982two} states and their lattice counterparts,  Fractional Chern Insulators (FCIs) \cite{parameswaran2013fractional}. In these systems, the fundamental low-energy quasiparticles carry fractional charge and lack an adiabatic connection to the physical electrons \cite{laughlin1983anomalous}. Nevertheless, theory predicts that bulk single-electron excitations of the FQH liquid remain well defined and long lived, though they occur only at high energies \cite{rezayi1987electron,he1993tunneling,Jain2005}. These theoretical predictions have recently gained renewed attention~\cite{yue2024electronic,gattu2025unlocking,pichler2025single,adhidewata2025excitation}, motivated in part by new scanning tunneling microscopy measurements of FCI spectral densities in graphene \cite{farahi2023broken} and  moir\'e materials \cite{hu2025high}. This progress raises a central question: if the single-particle Green's function primarily probes the high-energy sector of FQH physics, how much information can it reveal about the system's low-energy topological order?
 
\begin{figure*}[t!]
    \centering
    \includegraphics[width=0.9\textwidth]{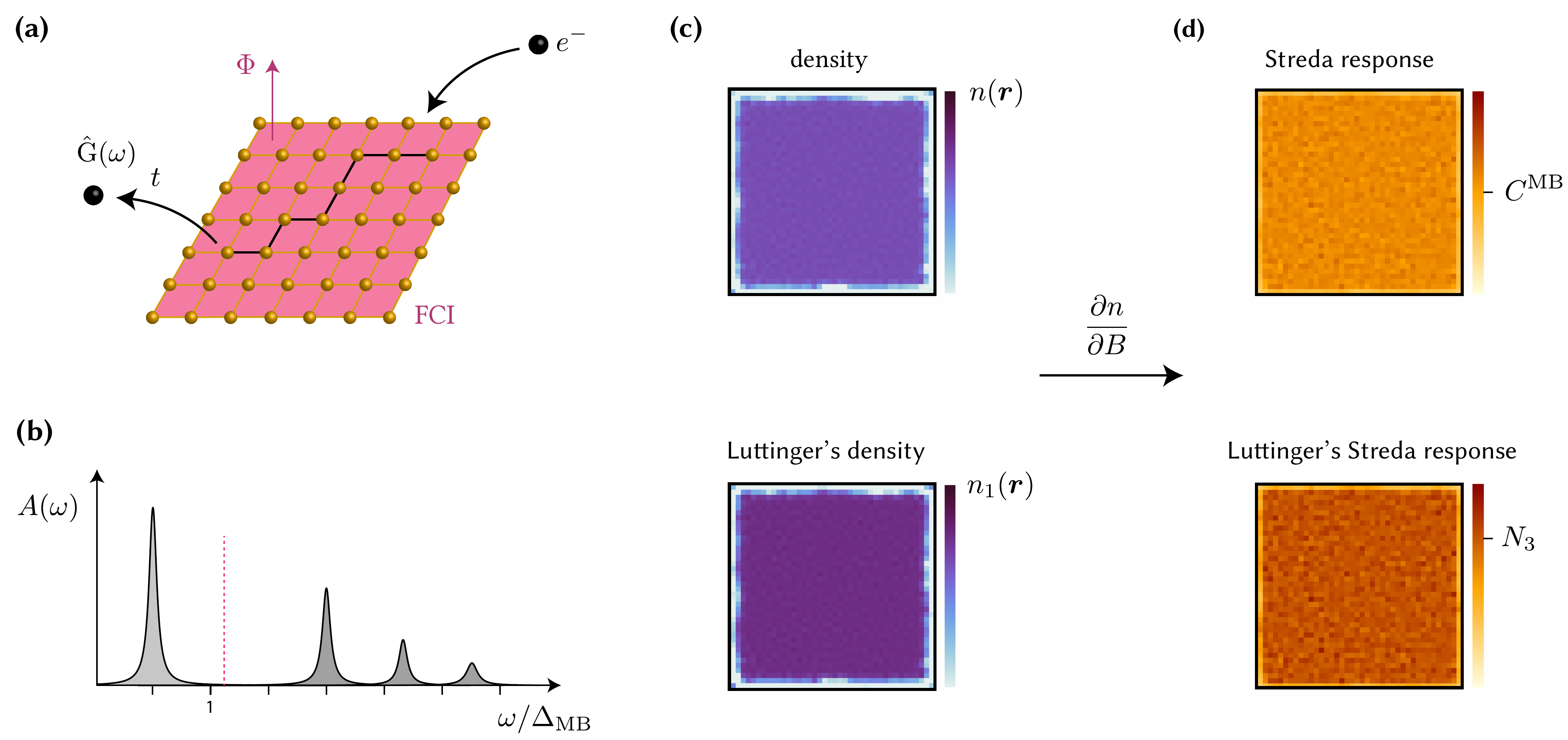}
    \caption{\textbf{Schematic illustration of the system, approach and key results.} \textbf{(a)} We study the bulk single-particle Green's function $\mathrm{G}(\omega)$ of interacting electrons in a uniform magnetic field realizing a fractional Chern insulator. \textbf{(b)} Schematic spectral function of a FQH state. The  electronic spectral density exhibits sharp coherent hole-like and electron-like peaks, corresponding to long-lived charged excitations located well above the neutral many-body gap $\Delta_{\mathrm{MB}}$. The dashed red line indicates the emergence of a zero of $\mathrm{G}$ within the charge gap. \textbf{(c)} In the FCI phase, the particle density and the Luttinger-count density differ, signaling a violation of Luttinger's theorem.  \textbf{(d)} The St\v{r}eda response of the particle density evaluated deep within the bulk yields the fractional many-body Chern number $C^{\mathrm{MB}}$, whereas the St\v{r}eda response of the Luttinger-count density $n_1$ produces an integer valued $N_3[\mathrm{G}]$ different from $C^{\mathrm{MB}}$.}
    \label{fig:fig1}
\end{figure*} 

Topological properties of non-interacting (and weakly-interacting) quantum matter are known to be encoded in the topology of their single-particle Green's functions~\cite{ishikawa1986magnetic,volovik2003universe,volovik2007quantum}. In particular, Ishikawa and Matsuyama~\cite{ishikawa1986magnetic} proposed that the quantized Hall conductivity of two-dimensional insulators can be expressed as $\sigma_{{\rm H}}\!=\! (e^2/h) N_3[\mathrm{G}]$, where $N_3[\mathrm{G}]$ denotes a topological invariant built from the single-particle Green's function $\mathrm{G}(\bm{k})$:

\be
	 N_3[\mathrm{G}]  =\frac{\epsilon^{\mu \nu \gamma}}{24 \pi^2}\!\!\int\!d^3k\; e^{k_0 0^{+}}\mathrm{Tr}\left(\mathrm{G}^{-1}\frac{\partial \mathrm{G}^{}}{\partial {k_\mu}}\mathrm{G}^{-1}\frac{\partial \mathrm{G}^{}}{\partial {k_\nu}}\mathrm{G}^{-1}\frac{\partial \mathrm{G}^{}}{\partial {k_\gamma}}\right).
\label{eq:N3}
\ee
Here $\mathrm{G}(\bm{k})$ denotes the Matsubara Green's function, $\bm{k} = (i\omega,k_x,k_y)$ is a three-vector in frequency-momentum space, $\epsilon^{\mu \nu \gamma}$ is the Levi-Civita symbol and the trace runs over all internal degrees of freedom. The initial derivation of Ref.~\cite{ishikawa1986magnetic} appeared to be non-perturbative, relying only on Ward-Takahashi identities~\cite{ward1950identity,takahashi1957generalized}, which led to the expectation that the relation $\sigma_{{\rm H}}\!=\! (e^2/h) N_3[\mathrm{G}]$ should persist beyond the weakly-interacting regime. However,  the integer nature of $N_3[\mathrm{G}]\!\in\!\mathbb{Z}$ suggests that the Ishikawa-Matsuyama relation should break down in the FQH regime, where the Hall conductivity is known to be fractionally quantized according to the many-body Chern number $\sigma_{{\rm H}}\!=\!(e^2/h) C^{\mathrm{MB}}$~\cite{niu1984quantised,avron1985quantization}. The possible multivaluedness of the Green's functions~\cite{ishikawa1986magnetic,wu2021anomalous} was proposed  as a way to reconcile the apparent discrepancy between the integer quantization of $N_3[\mathrm{G}]$ and the fractionally quantized values of the Chern number $C^{\mathrm{MB}}$ in the strongly-correlated FQH regime. 
In later years, the equivalence between $N_3[\mathrm{G}]$ and the many-body Chern number was cast into doubt. While $C^{\mathrm{MB}}$ cannot change without the collapse of the many-body gap,  $N_3[\mathrm{G}]$ can change when a zero of the single-particle Green's function crosses the chemical potential~\cite{gurarie2011single}. In fact, deviations between $N_3[\mathrm{G}]$ and $C^{\mathrm{MB}}$ were explicitly demonstrated, in both numerical and analytical studies~\cite{gurarie2013topological,he2016topological,zhao2023failure,peralta2023connecting,blason2023unified,bollmann2024topological}. For FQH states, the discrepancy was shown indirectly in Ref.~\cite{gurarie2013topological}, by invoking the bulk-edge correspondence~\cite{Essin2011}, and relying on the Green's functions of an effective low-energy edge theory~\cite{wen1990chiral}. More recently, Ref.~\cite{blason2023unified} clarified that the original Ishikawa-Matsuyama relation~\cite{ishikawa1986magnetic} implicitly relied on perturbative assumptions.


These developments raise fundamental questions about the physical significance of the $N_3[\mathrm{G}]$ invariant in strongly-interacting topological matter, both in systems lacking genuine topological order~\cite{he2016topological,zhao2023failure} and in topologically-ordered (FQH and FCI) phases~\cite{gurarie2013topological,gurarie2015topological}. A recent advance in this direction was made in Ref.~\cite{peralta2023connecting}, which reformulated $N_3[\mathrm{G}]$ within the framework of Luttinger's theorem \cite{luttinger1960ground} and the St\v{r}eda-Widom relation \cite{widom1982thermodynamic,Streda1982theory}. In Luttinger and Ward's original proof, \cite{luttinger1960ground} the total particle number $N$ is split into the Luttinger count $N_1[\mathrm{G}]$ and the Luttinger integral $\Delta N_1$. In a Fermi liquid, $N_1[\mathrm{G}]$ counts the number of single-particle states available to quasiparticles, whereas the Luttinger integral vanishes. Ref.~\cite{peralta2023connecting} applied an analogous decomposition to the St\v{r}eda-Widom relation \cite{widom1982thermodynamic,Streda1982theory}, which states that $C^{\mathrm{MB}} = \Phi_{0}\partial N/\partial \Phi$, with $\Phi_0 = hc/e$ the flux quantum and $\Phi$ the total flux threading the system. Within this viewpoint, $N_3$ captures the St\v{r}eda response of the Luttinger count $\Phi_0\partial N_1/\partial \Phi$ \footnote{This equality holds only in the absence of poles and zeros of the Green's function at the chemical potential; see \cite{peralta2023connecting} for details}. Thus, the many-body Chern number is equal to $N_3$ provided the Luttinger integral vanishes and Luttinger's theorem holds. The possible breakdown of Luttinger's theorem in strongly-correlated quantum matter has been intensely studied over the past decades~\cite{chubukov1996fermi,chubukov1997electronic,altshuler1998luttinger,georges2001quantum,rosch2007breakdown,heath2020necessary,Zitko2021,skolimowski2022luttinger,setty2024electronic,blason2023unified,farid2009comment,farid2012comment}, and the precise conditions for this violation have been a subject of intense debate~\cite{farid2009comment,farid2012comment}.  In particular, pioneering works already revealed non-zero values of the Luttinger integral $\Delta N_1$ in different settings, including antiferromagnetic metals~\cite{altshuler1998luttinger} and in the Sachdev–Ye–Kitaev model~\cite{georges2001quantum}. For the purpose of our work, a key insight from this literature is that Luttinger's theorem can fail when the single‑particle Green’s function develops zeros~\cite{chubukov1996fermi,chubukov1997electronic,altshuler1998luttinger,rosch2007breakdown,Zitko2021,skolimowski2022luttinger,setty2024electronic,blason2023unified,stanescu2006fermi,stanescu2007theory,dave2013absence,scheurer2018topological}.

Despite this progress, a direct evaluation of Green's-function-based topological invariants deep within the bulk of a fractionalized phase has remained elusive. This manuscript closes that gap by presenting a direct calculation of single-particle Green's functions in the bulk of an FCI state realized in the fermionic Harper-Hofstadter-Hubbard model~\cite{harper2015hofstadter,palm2023fractional}, obtained via exact diagonalization. We demonstrate the breakdown of Luttinger's theorem in this setting by explicitly evaluating $N_1$ and $\Delta N_1$, providing what is, to the best of our knowledge, the first exact numerical computation of these quantities in a strongly correlated topological phase. We then show -- both analytically and numerically -- that the fractional character of the many-body Chern number $C^{\mathrm{MB}}$  is encoded in the St\v{r}eda response of the Luttinger integral, $\Phi_0\partial (\Delta N_1)/\partial \Phi\!\in\!\mathbb{Q}$, rather than in the St\v{r}eda response of the Luttinger count, $\Phi_0\partial N_1/\partial \Phi\!=\!N_3\!\in\!\mathbb{Z}$. Thereafter, we show analytically that $N_3$ is fully determined by the Luttinger count $N_1$ together with the single-particle Chern numbers of the occupied Bloch bands, provided interaction-induced Bloch-band mixing is neglected. Finally, we propose a protocol for extracting the Luttinger count, the Luttinger integral and their St\v{r}eda response from the probes of the local density of states, now available experimentally for FCI states \cite{farahi2023broken,hu2025high}. A schematic summary of our approach and main results is shown in \cref{fig:fig1}.

The manuscript is organized as follows. In \cref{sec:Luttinger's theorem}, we review the relation between topological invariants of the single-particle Green's function and  Luttinger's theorem. In \cref{sec:Results}, we characterize the bulk single-particle spectral properties of the FCI, identifying the emergence of poles and zeros of the Green's function; we then present numerical evidence for the violation of Luttinger's theorem, and obtain the individual St\v{r}eda responses of the bulk particle density and its Luttinger constituents (i.e.~the local Luttinger's count and integral). In \cref{sec:N3 analytics} we provide an analytical framework that rationalizes these numerical findings, deriving an explicit expression for $N_3$ in the absence of Bloch band mixing. In \cref{sec:scheme}, we propose a practical scheme for extracting the Luttinger's count and integral, and their St\v{r}eda responses, from local density of states measurements. We conclude in \cref{sec:conclusions} with a discussion of the implications of our findings and possible future research directions.

\section{Particle Counting, Luttinger's Theorem and Topological Winding Numbers of the Green's Function}
\label{sec:Luttinger's theorem}

The particle number of an interacting system at a given chemical potential $\mu$ and zero temperature can be expressed in terms of the single-particle Green's function as
\begin{equation}
\label{eq:N_interacting}
N(\mu) = \frac{1}{2\pi i}\int dz e^{z0^+} \mathrm{Tr}[\mathrm{G}(z)] = \int_{-\infty}^{\mu} d\omega A(\omega),
\end{equation}
where $\mathrm{G}(z)$ is defined on the complex frequency plane and the integral in the first equality is understood as a principal value taken along a line parallel to the imaginary axis, $z=\mu+i\omega'$. The second equality is obtained by deforming the contour toward the real axis, $z=\omega\mp i 0^{+}$, which naturally defines the total spectral density or density of states
\begin{equation}
A(\omega) = \sum_\alpha A_\alpha(\omega) = \frac{1}{\pi} \sum_\alpha \Im \mathrm{G}_{\alpha\alpha}(\omega-i0^+),
\label{eq:A_omega}
\end{equation}
with $A_{\alpha}(\omega)$ the contribution from state $\alpha$. 

Following Luttinger and Ward~\cite{luttinger1960ground}, Eq.~\eqref{eq:N_interacting} can be decomposed into two terms: the Luttinger count $N_1[\mathrm{G}]$, which depends solely on the analytical structure of the propagator, and the Luttinger integral $\Delta N_1$, which explicitly involves the self-energy $\Sigma(z)$,
\begin{subequations}
\begin{align}
N &= N_1[\mathrm{G}] + \Delta N_1,\\
\label{N1}
N_1[\mathrm{G}] &= -\frac{1}{2\pi i}\int dz e^{z0^{+}}\mathrm{Tr}\left[\mathrm{G}^{-1}(z)\partial_z \mathrm{G}(z)\right],\\
\label{DeltaN1}
\Delta N_1&= \frac{1}{2\pi i}\int dz e^{z0^{+}}
\mathrm{Tr}\left[\mathrm{G}(z)\partial_z \Sigma(z)\right].
\end{align}
\label{eq:n1_n2}
\end{subequations}

\begin{figure}[t]
    \centering
    \includegraphics[width=0.9\columnwidth]{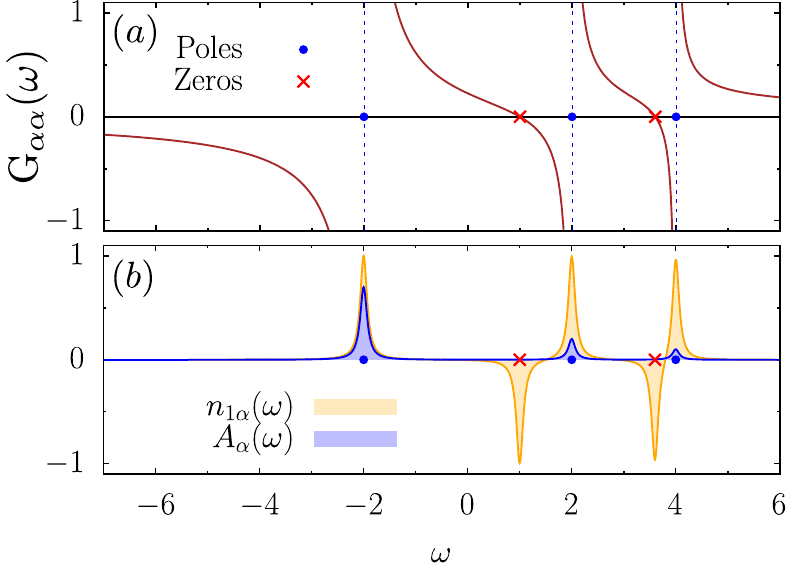}
    \caption{$(a)$ Schematic illustration of the diagonal element of a toy single-particle Green's function on the real-frequency axis, with poles at $\varepsilon_\alpha^{(p)} = -2, 2, 4$ (blue circles) and corresponding spectral weights $Z_\alpha^{(p)}=0.7, 0.2, 0.1$. The positions of the zeros $\chi_\alpha^{(s)}$ of $\mathrm{G}_{\alpha\alpha}(\omega)$ are indicated by red crosses. $(b)$ Spectral density $A_{\alpha}(\omega)$  [Eq.~\eqref{eq:A_omega}] and Luttinger counting kernel $n_{1\alpha}(\omega)$ [Eq.~\eqref{eq:n1_omega}].}
    \label{fig:fig2}
\end{figure}

 Luttinger's theorem holds whenever $\Delta N_1=0$, in which case the particle number is entirely determined by $N_1[\mathrm{G}]$~\cite{luttinger1960ground}. Crucially, when $\mathrm{G}(z)$ has no poles or zeros at the chemical potential, the contour in Eq.~\eqref{N1} can be closed in the lower half-plane, turning the Luttinger count into a bona-fide topological invariant: the first-order winding number of the Green's function~\cite{seki2017topological,gurarie2011single}. To clarify what this topological count actually evaluates in an interacting system, it is useful to make explicit the analytical structure of $\mathrm{G}(z)$. For a finite system, each diagonal element has the rational form~\cite{seki2017topological},
\begin{equation}
\label{eq:G_rational}
    \mathrm{G}_{\alpha\alpha}(z) = \sum_{p} \frac{Z_\alpha^{(p)}}{z-\varepsilon_\alpha^{(p)}} = \frac{\prod_{s}(z-\chi_\alpha^{(s)})}{\prod_p (z-\varepsilon_{\alpha}^{(p)})},
\end{equation}
where $Z_\alpha^{(p)}>0$ are the Lehmann spectral weights, and $\varepsilon_\alpha^{(p)}$ and $\chi_\alpha^{(s)}$ denote the poles and zeros of $\mathrm{G}_{\alpha\alpha}(z)$, respectively. 
In this structure, each zero necessarily lies between two consecutive poles along the real-frequency axis, reflecting the redistribution of spectral weight induced by interactions. Figure~\ref{fig:fig2}(a) illustrates this behavior schematically. Choosing $\alpha$ to label the basis that diagonalizes $\mathrm{G}(z)$, the Luttinger count becomes
\begin{equation}
     N_1[G] = \sum_{\alpha}\int_{-\infty}^{\mu}d\omega \; n_{1\alpha}(\omega),
\end{equation}
with the Luttinger counting kernel
\begin{equation}
n_{1\alpha}(\omega) = \sum_p \delta(\omega-\varepsilon_\alpha^{(p)}) - \sum_s \delta(\omega-\chi_\alpha^{(s)}),
\label{eq:n1_omega}
\end{equation}
so that each pole below $\mu$ contributes $+1$, while each zero contributes $-1$. Because these contributions are discrete, the only way to modify this quantized invariant is for a pole or a zero to cross the chemical potential. Figure~\ref{fig:fig2}(b) compares $A_{\alpha}(\omega)$ and $n_{1\alpha}(\omega)$ for the toy Green's function of Fig.~\ref{fig:fig2}$(a)$. Taking, for instance, $\mu=0$, one immediately sees how Luttinger’s theorem can fail in a correlated insulator even in the absence of Green's functions zeros below the chemical potential: the spectral density $A_{\alpha}(\omega)$ integrates to the actual spectral weight of the peak below $\mu$, while the Luttinger count assigns that peak a unit weight. It is also important to note that, if the chemical potential is allowed to vary within the single-particle gap, the Luttinger count value changes by $\pm1$ when a Green's function zero crosses the Fermi level \cite{rosch2007breakdown}.

In the absence of poles and zeros at the chemical potential, Ref.~\cite{peralta2023connecting} further showed that the Ishikawa-Matsuyama invariant $N_3[\mathrm{G}]$ can be accessed directly by evaluating the St\v{r}eda response of the Luttinger count $N_1[\mathrm{G}]$:
\begin{equation}
    N_3[\mathrm{G}] = \Phi_0 \frac{\partial N_{1}[\mathrm{G}]}{\partial \Phi}\Bigg\rvert_{\mu},
    \label{N3_Streda}
\end{equation}
i.e., by monitoring how the Luttinger count responds to an infinitesimal flux $\Phi $ threading the system at fixed chemical potential $\mu$. This relation circumvents the need to evaluate the full three-form defining $N_3[\mathrm{G}]$, allowing one to work entirely at the level of the one-form Luttinger sum rule. 
When Luttinger's theorem is violated ($\Delta N_1 \neq 0$),
the discrepancy between the many-body Chern number and the Green's function based-invariant is then captured by the flux derivative of the Luttinger integral,
\begin{equation}
    \Delta N_3 = C^{\mathrm{MB}}-N_3[\mathrm{G}]= \Phi_0 \frac{\partial \Delta N_1}{\partial \Phi}\Bigg\rvert_{\mu}.
    \label{DeltaN3_Streda}
\end{equation}

\begin{figure*}[t!]
    \centering
    \includegraphics[width=\linewidth]{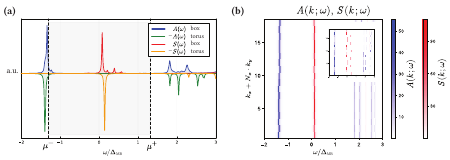}
    \caption{\textbf{Analytical structure of the Green's functions and the self-energy.} \textbf{(a)} Local spectral densities of the Green's function, $A(\omega)$, and the self-energy, $S(\omega)$, defined in \cref{eq:spectral} computed in the finite box (positive sign) and on the torus (negative sign). Dashed lines denote the electron addition and removal energies, $\mu^{+}=E_0(N+1)-E_0(N)$ and $\mu^{-} = E_0(N) - E_0(N-1)$, extracted from the torus geometry, where $E^{N\pm1}_0$ are the ground state energies in the $N\pm1$-particle sectors. Note that $A(\omega)$ and $S(\omega)$ are plotted in different scales. \textbf{(b)}  Momentum-resolved spectral densities $A(k,\omega)$ and $S(k,\omega)$ on the torus geometry. The main panel shows spectra obtained from the thermodynamically averaged Green's function over the degenerate ground-state manifold [\cref{eq:therm_GF}], while the inset presents the result of averaging only over a single ground-state [\cref{eq:signgle_gs_GF}]. The frequency is measured in units of the fixed-$N$ many-body gap $\Delta_{\mathrm{MB}}$ computed on the torus geometry, with zero frequency set at $(\mu^{+}+\mu^{-})/2$. The momenta $k_x$ and $k_y$ are measured in units of $\frac{2\pi}{N_x}$ and $\frac{2\pi}{q N_y}$}
    \label{fig:fig3}
\end{figure*}

In what follows, we exploit this connection by numerically evaluating $N_1[\mathrm{G}]$, $\Delta N_1$, and their flux derivatives in a fractional Chern insulator, enabling a direct diagnosis of both the breakdown of Luttinger's theorem and an exact evaluation of these Green's function-based topological invariants in a topologically ordered system from exact diagonalization calculations.

\section{Green's function and  topological invariants from exact diagonalization}
\label{sec:Results}

In view of performing explicit numerical calculations, we henceforth consider the Harper-Hofstadter-Hubbard model, paradigmatic model for FCI states \cite{harper2015hofstadter,palm2023fractional}. The model describes a system of $N$ interacting spinless fermions subjected to a uniform magnetic field with the Hamiltonian:  
\begin{equation}
\label{eq:Hamiltonian}
\begin{split}
H = \sum_{n,m} \left[ 
  -t \left( 
    c^{\dagger}_{n,m+1} c_{n,m} 
    + c^{\dagger}_{n+1,m} c_{n,m} e^{i 2\pi m \varphi} 
    + h.c. 
  \right) \right.\\
  \left.
  + V\hat{P}\! :\! n_{n,m} \left( n_{n\pm 1,m} + n_{n,m\pm 1}  \right)\! :\! \hat{P}
\right],
\end{split}
\end{equation}
where $c^{(\dagger)}_{n,m+1}$ annihilates (creates) a fermion on the lattice site $(n,m)$, $\varphi$ is the magnetic flux per plaquette in units of the flux quantum, $V$ is the nearest neighbor interaction strength, $:$ stands for the normal ordering and $\hat{P}$ projects the dynamics onto the lowest Hofstadter band (LHB). In the weak magnetic field regime $\varphi\ll1$, the LHB continuously connects to the lowest Landau Level. At fractional fillings $\nu = N/N_{\Phi}$, where $N_{\Phi}$ is the number of flux quanta piercing the system, the projected interacting model supports fractional Chern insulator phases, that closely resemble the corresponding Laughlin states in the continuum \cite{scaffidi2012adiabatic,parameswaran2013fractional}. 

To investigate the fate of the Luttinger's theorem and the Ishikawa-Matsuyama invariant in these fractional Chern insulator states, we proceed in two complementary ways. First, we study the system in a finite box with open boundary conditions and compute the single-particle Green's function $\mathrm{G}(\bm{r},\bm{r'};z)$ and self-energy $\Sigma(\bm{r},\bm{r'};z)$ using Lanczos exact diagonalization (see \cref{sec:ap_numerics} for details). This approach allows us to evaluate spatially resolved versions of the Luttinger count and integral in Eq.~\eqref{eq:n1_n2},

\begin{equation}
\label{eq:localn1n2}
\begin{split}
    n_1(\bm{r}) &= -\frac{1}{2\pi i}\int_{\mathcal{C}} dz \sum_{\bm{r'}}\;  \mathrm{G}^{-1}(\bm{r},\bm{r'};z)\partial_z \mathrm{G}(\bm{r'},\bm{r};z)\\
    \Delta n_1(\bm{r}) &= \frac{1}{2\pi i}\int_{\mathcal{C}} dz \sum_{\bm{r'}}\; \partial_z \Sigma(\bm{r},\bm{r'};z) \mathrm{G}(\bm{r'},\bm{r};z) , \\
\end{split} 
\end{equation}
where the contour $\mathcal{C}$ 
encloses all occupied poles in the complex frequency plane. In the bulk region, the insulating gap ensures that both $G$ and $\Sigma$ decay exponentially with $|\bm{r}-\bm{r'}|$, according to a finite correlation length, see Appendix \ref{sec:app:finite_size}. Evaluating these quantities deep within the bulk region therefore provides a faithful representation of the intrinsic bulk response to a magnetic perturbation, see Eqs.~\eqref{N3_Streda} and~\eqref{DeltaN3_Streda}.

As a second, complementary approach, we apply the same exact diagonalization method to the Hamiltonian defined on a torus. While the St\v{r}eda response is hard to probe in this geometry due to the Dirac quantization of the magnetic flux, the torus preserves translational invariance and eliminates edge effects, allowing us to benchmark the results extracted from the bulk of the open-boundary box. It also enables investigation of the momentum dependence of $\mathrm{G}(\bm{k};z)$ and $\Sigma(\bm{k};z)$, a key aspect for elucidating the behavior of the winding number $N_3[G]$. 

We focus on the most stable fractional Chern insulating phase for fermions, which occurs at filling $\nu\smeq\frac{1}{3}$. To realize this state, we consider a magnetic flux per plaquette $\varphi=\frac{1}{5}$, corresponding to a nearly flat LHB that closely resembles the lowest Landau level. For the simulations, we consider $N_p = 6$ particles on a finite open-boundary box of $N_x\times N_y = 11\times10$ sites, while the torus geometry has $N_x\times N_y = 9\times10$ sites, covering the same total area. The interaction strength is set to $V/t=10$. This choice of parameters stabilizes the same phase in both geometries as we demonstrate below.

\subsection{Bulk Green's Functions of a Laughlin-type FCI state}
\label{sec:Results_2}

 We begin by characterizing the bulk properties of the system through the frequency dependence of the retarded Green's function and the self-energy. To this end, we focus on the corresponding local spectral densities,

\be
\label{eq:spectral}
\begin{split}
A(\omega)&= -\frac{1}{\pi} \Im \left[ \mathrm{G}(\bm{r}_0,\bm{r}_0;\omega+i0^{+})\right]\\
S(\omega)&= -\frac{1}{\pi} \Im \left[ \Sigma(\bm{r}_0,\bm{r}_0;\omega+i0^{+})\right],   
\end{split}
\ee
with the site $\mathbf{r}_0$ chosen at the center of the sample. In the bulk, the residual dependence of $A(\omega)$ and $S(\omega)$ on the spatial coordinate $\textbf{r}_0$ arises solely due to finite-size effects, and can therefore be neglected \footnote{The explicit lattice translational symmetry is broken in the presence of the homogeneous magnetic field. All the gauge invariant quantities, however have to preserve the translational symmetry. The elements of the Green's functions and the self-energy which are diagonal in space indices are gauge invariant and therefore are translational invariant \cite{kita2005theory}}. 
However, in a small finite-size box, the self-energy spectral density becomes particularly sensitive to edge effects, even though the correlation length for our parameter set is only on the order of a single lattice constant; see \cref{sec:app:finite_size}. In this sense, a comparison with the results obtained on the torus geometry are relevant.

The spectral densities $A(\omega)$ and $S(\omega)$ provide complementary information: $A(\omega)$ resolves the poles of the Green's function and thus the single-particle density of states, whereas $S(\omega)$ captures the location of its zeros. The resulting spectra are shown in \cref{fig:fig3}. Panel $\textbf{(a)}$ displays $A(\omega)$ and $S(\omega)$ for both box and torus geometries as a function of frequency $\omega$, with the torus data plotted with opposite sign for ease of comparison. Interactions split the quasi-flat single-particle LHB, opening a clear charge gap that separates the density of states corresponding to electron removal (occupied states) from that corresponding to electron addition (empty states). Crucially, this gap is accompanied with the emergence of a single zero of the Green's function (or equivalently, a pole in the self-energy), visible as a peak in $S(\omega)$. This pole-zero structure reflects a strong redistribution of spectral weight induced by the correlations. 

\begin{figure*}[t!]
    \centering
    \includegraphics[width=\linewidth]{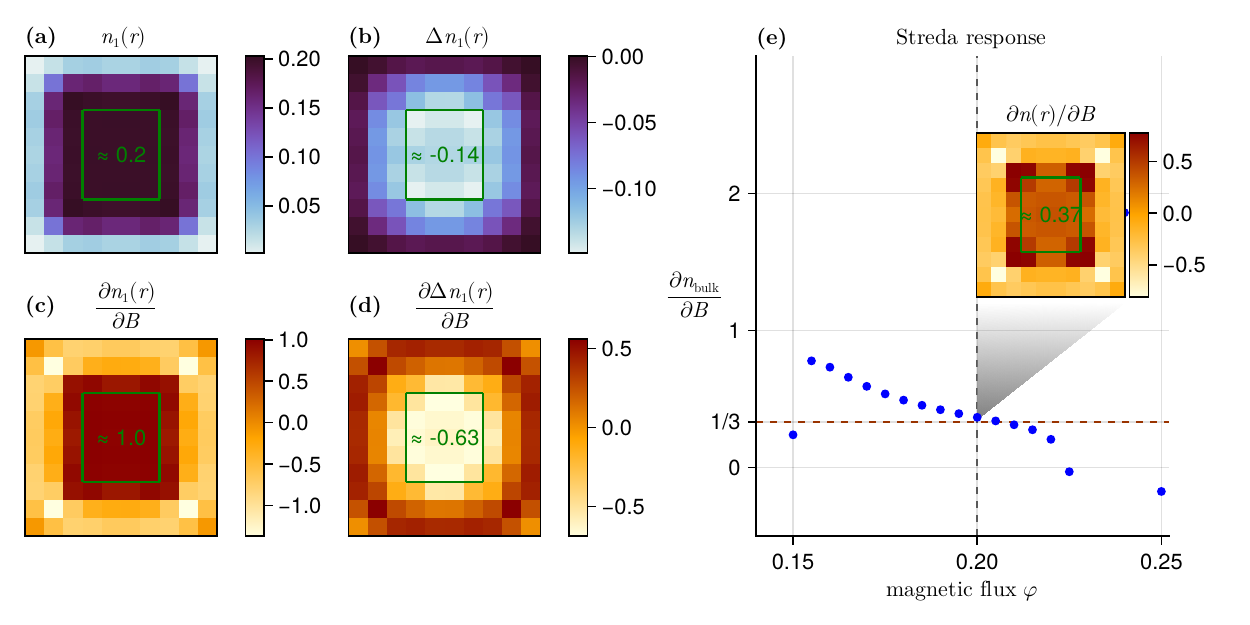}
    \caption{\textbf{St\v{r}eda-response and Luttinger's theorem violation in the FCI state.} \textbf{(a,b)} Spatial distributions of the local Luttinger count $n_1(\bm{r})$ and Luttinger integral $\Delta n_1(\bm{r})$, defined in \cref{eq:localn1n2}.  The presence of a finite, negative $\Delta n_1(\bm{r})$ in the bulk directly signals the breakdown of Luttinger's theorem. \textbf{(c,d)} St\v{r}eda responses of the Luttinger count and the Luttinger integral densities. The derivative of $n_1$ with respect to magnetic field (averaged over the bulk sites enclosed by the green contour) yields an integer-valued response, corresponding to an Ishikawa-Matsuyama invariant $N_3[\mathrm{G}] \simeq 1$. By contrast, the St\v{r}eda response of $\Delta n_1$ is fractional and accounts for the difference between $N_3[\mathrm{G}]$ and the many-body Chern number. \textbf{(e)} St\v{r}eda response of the total bulk electron density as a function of magnetic flux per plaquette. The developing plateau near $1/3$ is consistent with the fractional Hall conductance of the FCI state realized at flux $\varphi = 1/5$. The inset shows the spatially-resolved magnetic response of the total density $n(\bm{r})$ at $\varphi = 1/5$.}
    \label{fig:fig4}
\end{figure*}

Remarkably, the spectral function $A(\omega)$ exhibits sharp, coherent features: a dominant hole-like peak and several pronounced electron-like peaks, superimposed on a weaker incoherent background, indicating the presence of well-defined, long-lived high-energy quasiparticle excitations. The overall structure, and its pronounced particle-hole asymmetry, closely mirror previous results obtained for FQH states on the sphere~\cite{Rezayi1987,Jain2005,Vignale2006,Gattu2024,yue2024electronic}. In particular, the strong asymmetry between the hole and particle sectors, first noticed in Ref.~\cite{Rezayi1987}, is a hallmark of Laughlin-type fractionalized states and reflects the fundamentally different nature of quasiholes and quasielectrons~\cite{girvin1984particle}. Physically, the simplicity of the hole sector for this particular filling reflects the fact that for a $\frac{1}{m}$-Laughlin state removing an electron is exactly equivalent to inserting $m$ quasi-hole excitations~\cite{girvin1984particle}. The resulting state is therefore very close to an eigenstate of the many-body Hamiltonian, which leads to a single coherent peak on the hole side~\cite{Rezayi1987}. In contrast, an inserted electron decomposes into multiple excitations that cannot be represented by a unique low-energy eigenstate. This fragmented particle-side spectrum, with a small number of dominant coherent peaks, has been understood within the composite-fermion framework~\cite{Jain2005,Gattu2024,yue2024electronic}. 
Notably, in our bulk box calculations we observe two pronounced particle-like peaks, in close correspondence with the results for $\nu=1/3$ in Refs.~\cite{Jain2005,Gattu2024,yue2024electronic}. On the torus, the particle side of the spectrum is more involved as first noted in Ref.~\cite{adhidewata2025excitation}. However, the difference between the torus and the box results can be expected to disappear in the thermodynamical limit (see \cref{sec:app:finite_size}). 

The main panel in Fig.~\ref{fig:fig3}\textbf{(b)} shows the momentum-resolved spectral densities $A(k,\omega)=-\frac{1}{\pi} \Im \left[ \mathrm{G}(k,\omega+i0^{+})\right]$ and $S(k,\omega)= -\frac{1}{\pi} \Im \left[ \Sigma(k,\omega+i0^{+})\right]$, plotted as a function of the real frequency $\omega$ and magnetic crystal momentum $k$ on the torus geometry. The Green's function is obtained from a thermodynamic average over the three-fold ground-state manifold $\{|\Psi_0^j\rangle,{j=1,2,3}\}$ , such that

\be
\label{eq:therm_GF}
\begin{split}
    A(k,\omega) = \frac{1}{3} \sum_{j=1}^3\sum_n  \bigl[ & |\langle \Psi_0^j|\hat{c}_k|\Psi_n\rangle|^2\delta(\omega -E_n + E_0) + \bigr.\\
    \bigl. & |\langle \Psi_0^j|\hat{c}_k^\dagger|\Psi_n\rangle|^2\delta(\omega + E_n - E_0) \bigr] ,
\end{split}
\ee
with $|\Psi^j_n\rangle$ the eigenstates of the Hamiltonian in the ${N\pm1}$ particle sector and $E_n$ their corresponding eigenvalues. 
Both spectral functions display only a weak dependence on momentum. We note that in the continuum FQH limit, the Green's function is expected to be strictly momentum independent~\cite{haussmann1996correlation,currienon}. This striking feature of the Green's functions results from the homogeneity of the FQH liquid and the flatness of the Landau levels \cite{currienon}. Accordingly, the residual momentum dependence reflects the finite dispersion of the LHB at flux $\varphi=1/5$ and is expected to diminish as the system approaches the continuum limit. The role of the ground-state averaging is illustrated in the inset of Fig.~\ref{fig:fig3}\textbf{(b)}, which shows the spectral densities obtained by restricting the averaging to a single ground-state $|\Psi_0^{1}\rangle$:

\be
\label{eq:signgle_gs_GF}
\begin{split}
 \tilde{A}(k,\omega) = \sum_n  \bigl[ & |\langle \Psi_0^1|\hat{c}_k|\Psi_n\rangle|^2\delta(\omega -E_n + E_0) + \bigr.\\
    \bigl. & |\langle \Psi_0^1|\hat{c}^\dagger_k|\Psi_n\rangle|^2\delta(\omega + E_n - E_0) \bigr].
\end{split}
\ee
In this case, a pronounced momentum dependent structure appears in the positions of the poles of both $\tilde{S}(k,\omega)$ and $\tilde{A}(k,\omega)$, repeated with period $2\pi/N_x\cdot3$. This periodicity can be naturally traced back to the different sets of momenta available for the excited states in the thin-torus limit \cite{tao1983fractional,bergholtz2008quantum}. While the actual eigenstates in two-dimensions are different from those in the quasi-one dimensional limit, this characteristic momentum periodicity is expected to persist for any finite circumference of the torus. A more detailed discussion of this behavior is provided in \cref{sec:ap single gs}. Our results are consistent with the momentum dependence of the local density-of-states reported in the thin cylinder limit in a recent preprint \cite{adhidewata2025excitation}.

\subsection{Numerical Study of Green's Function Topological Invariants}
\label{sec:Results_3}

Having characterized the poles and zeros of the Green's function along the real-frequency axis, we now turn to the central result of this work: a direct numerical evaluation of Green's function topological invariants in a fractional Chern insulator. Specifically, we compute the densities of the Luttinger count $n_1(\bm{r})$ and the Luttinger integral $\Delta n_1(\bm{r})$, together with their magnetic-field responses, by numerically performing the contour integration in the complex frequency plane prescribed by Eq.~\eqref{eq:localn1n2}. This procedure provides an explicit, non-perturbative determination of $N_1[\mathrm{G}]$ and $\Delta N_1$, and their St\v{r}eda responses, $N_3[\mathrm{G}]$ and $\Delta N_3$, in a strongly correlated, topologically ordered state. The calculations are carried out in the box geometry, allowing direct access to the bulk Green's functions.

The position of the chemical potential within the charge gap, relative to the zero of the Green’s function, determines the value of the Luttinger count, as anticipated in the general overview of \cref{sec:Luttinger's theorem}. Topological insulators are special in this respect:~In contrast to a topologically trivial insulator -- where the chemical potential can be fixed by simply taking the zero-temperature limit \cite{farid2012comment} -- a topological insulator with open boundaries responds differently:~shifting the chemical potential alters the occupation of the conducting edge states and thus represents a physically meaningful operation, one that can be tuned experimentally through electrostatic gating. In this section, we fix the chemical potential between the main electron-removal peak and the in-gap self-energy pole. This choice allows us to evaluate the Luttinger count and integral directly from the ED data while avoiding spurious finite-size effects. For other choices of $\mu$, the contour integration in the complex frequency plane becomes more sensitive to finite-size edge contributions, as discussed in detail in \cref{sec:ap_countour}. The integration can however be performed when the finite-size effects stemming from the edge poles of the Green's functions are eliminated by hand. We follow that route in \cref{sec:scheme}.

The spatial distributions of Luttinger count density $n_1(\bm{r})$ and the Luttinger integral density $\Delta n_1(\bm{r})$ are shown in \cref{fig:fig4} \textbf{(a)} and \textbf{(b)}, respectively. Both quantities exhibit well-defined plateaus in the bulk of the system, indicating that finite-size and boundary effects are well controlled. Their bulk-averaged values unambiguously reveal a violation of Luttinger's theorem in the FCI state. The bulk value of the Luttinger count, $n_1\simeq 0.2$, corresponds to a completely filled LHB, since the magnetic unit cell contains five lattice sites. This value of the Luttinger count is in agreement with the analytical structure of the Green's function shown in \cref{fig:fig3}, where a single pole appears below the chosen chemical potential for each LHB state. By construction, the Luttinger count assigns unit weight to each such pole, thereby overestimating the  physical particle density, as anticipated in \cref{sec:Luttinger's theorem}. The Luttinger integral $\Delta n_1$ must therefore be negative,  compensating for the difference between the actual spectral weight of the hole peak and unity, as illustrated in~\cref{fig:fig2}~\textbf{(b)}. 

The St\v{r}eda responses of the local electronic density $n(\bm{r})$ and of its constituents, the Luttinger count $n_1(\bm{r})$ and the Luttinger integral densities $\Delta n_1(\bm{r})$, are shown in Figs.~\ref{fig:fig4}~\textbf{(c--e)}. Resolving the total density response of individual Luttinger components reveals a striking separation of integer and fractional contributions. The bulk St\v{r}eda response of the Luttinger count, $\partial n_1/\partial B $, displayed in Fig.~\ref{fig:fig4}\textbf{(c)}, yields a quantized value $N_3[\mathrm{G}]\simeq 1$, indicating that the Ishikawa-Matsuyama invariant remains strictly integer in an FCI state in spite of the fractional value of the many-body Chern number. In contrast, the fractional Hall response is entirely captured by the St\v{r}eda response of the Luttinger integral, $\partial \Delta n_1/\partial B\!\approx\!-2/3$, shown in Fig.~\ref{fig:fig4}\textbf{(d)}. The associated invariant $\Delta N_3$ precisely accounts for the difference between $N_3[\mathrm{G}]$ and the fractional many-body Chern number $C_{\mathrm{MB}}=1/3$, consistent with the theoretically predictions of Ref.~\cite{peralta2023connecting}. The many-body Chern number is directly reflected in the St\v{r}eda response of the total bulk density, shown in Fig.~\ref{fig:fig4}\textbf{(e)} as a function of the flux per plaquette $\varphi$, where a clear developing plateau near $1/3$ emerges around $\varphi=1/5$.

Our results appear to contradict previous analytical predictions for the value of $N_3[\mathrm{G}]$ in FQH states~\cite{gurarie2013topological}, where a $\mu$-independent value of $N_3[\mathrm{G}]\!=\!3$  was inferred for the $\nu\!=\!1/3$ Laughlin state by evaluating a momentum-resolved Luttinger count. Importantly, that previous approach used the low-energy form of the electron edge Green's function~\cite{wen1990chiral} and invoked the bulk-edge correspondence proven in Ref.~\cite{Essin2011}. We argue that this discrepancy arises because the ``edge winding number" entering the bulk-edge correspondence in Ref.~\cite{Essin2011} is defined in terms of the full electron edge Green's function over all frequencies, whereas the low-energy edge theory constrains only its behavior arbitrarily close to the chemical potential. In an FCI state, the electron edge Green's function generically contains non-universal high-energy structure that is invisible to the low-energy theory but can nevertheless contribute to the Green's function winding. We therefore conjecture that reconstructing $N_3[\mathrm{G}]$ from the low-energy electron edge data alone is not generically valid in fractionalized phases.

Summarizing, our numerical studies demonstrate that the Ishikawa-Matsuyama invariant takes the value $N_3[\mathrm{G}]\!=\!1$ deep in the bulk of a Laughlin-type FCI state, for a given choice of the chemical potential in the gap. In the following Section, we will analytically demonstrate that this topological invariant is fully determined by the Luttinger count $N_1$ together with the single-particle Chern number of the occupied Bloch band, provided interaction-induced Bloch-band mixing is neglected.

\section{Analytical Evaluation of $N_3[\mathrm{G}]$ in the Absence of Bloch Band Mixing}
\label{sec:N3 analytics}

To clarify the origin of the numerical observations above and their deviation from the predictions of Ref.~\cite{gurarie2013topological}, we analytically evaluate the Green's function winding number $N_3[\mathrm{G}]$ in the limit where Bloch band mixing can be neglected. By this we mean that, throughout the frequency range relevant for determining $N_3[\mathrm{G}]$, the interacting Green's function is well approximated as diagonal in the band index. This approximation is justified whenever the interaction scale is small compared to the energy gaps separating different Bloch bands, so that interaction-induced virtual transitions into other bands only weakly affect the Green's function in the frequency range relevant for the contour integration defining the invariant. It is also exact whenever the many-body Hamiltonian has been explicitly projected onto a chosen band, in which case band mixing is absent by construction, as in Eq.~\eqref{eq:Hamiltonian}.

We begin by expressing the interacting one-body Green's function in the basis of single-particle magnetic Bloch orbitals $|u_{\alpha\bm{k}}\rangle = \hat{c}^{\dagger}_{\alpha\bm{k}}|0\rangle$. Magnetic translation symmetry implies that crystal momentum remains a good quantum number, so that the Green's function decomposes into independent $\bm{k}$ sectors. Within each sector, and assuming the propagator is diagonal in the band index, we may write
\begin{equation}
\hat{\mathrm{G}}_{\bm{k}}(z) = \sum_{\alpha} g_{\alpha\bm{k}}(z)\,
|u_{\alpha\bm{k}}\rangle\langle u_{\alpha\bm{k}}| .
\label{ansatz}
\end{equation}
Here we represent the Green's function as an operator acting on the single-particle Hilbert space spanned by the magnetic Bloch orbitals. This makes the momentum derivatives appearing in Eq.~\eqref{eq:N3} explicit, since they act both on the scalar functions $g_{\alpha\bm{k}}(z)$ and on the $\bm{k}$-dependence of the basis states. Alternatively, one may introduce a $\bm{k}$-independent reference basis $\{|n\rangle\}$ and express the propagator in matrix form,
\begin{equation}
 \mathrm{G}_{\bm{k}}^{}(z)=U_{\bm{k}}g_{\bm{k}}(z)U^{\dagger}_{\bm{k}}, 
 \label{ansatz_matrix}
\end{equation}
where $g_{\bm{k}}(z) = \mathrm{diag}(g_{\alpha\bm{k}}(z))$ and $U_{\bm{k}}^{n\alpha}=\langle n|u_{\alpha\bm{k}}\rangle$ is a unitary transformation encoding the $\bm{k}$-dependence of the Bloch basis.

At zero temperature, the scalar functions $g_{\alpha\bm{k}}(z)$ entering Eq.~\eqref{ansatz} are defined as
\begin{equation}
g_{\alpha\bm{k}}(z) = -\frac{1}{d_g}\sum_{j=1}^{d_g}
\int_{0}^{\infty} d\tau\, e^{z\tau}
\langle \Psi_0^{j}|
\mathcal{T}_{\tau}
\hat{c}^{}_{\alpha\bm{k}}(\tau)
\hat{c}^{\dagger}_{\alpha\bm{k}}(0)
|\Psi_0^{j}\rangle ,
\end{equation}
where $\mathcal{T}_{\tau}$ denotes imaginary-time ordering and
$|\Psi_0^{j}\rangle$ are the $d_g$ (topologically) degenerate many-body ground states. This expression corresponds to the zero-temperature limit of the finite-temperature Green's function in the presence of ground-state degeneracy, resulting in a statistical average over the degenerate manifold.

Substituting Eq.~\eqref{ansatz} (or equivalently the matrix representation 
in Eq.~\eqref{ansatz_matrix}) into the definition of the topological invariant Eq.~\eqref{eq:N3}, and using the fact that the diagonal elements $g_{\alpha\bm{k}}(z)$ possess a meromorphic structure in frequency (Eq.~\eqref{eq:G_rational}), one may perform the frequency integration explicitly, see \cref{sec:ap derivation N3_ansatz} for details. Assuming that no poles nor zeros of the propagator cross the chemical potential, one obtains

\begin{eqnarray}
\notag
    N_3[\mathrm{G}]&=&\frac{1}{2\pi}\sum_\alpha \int d^2k    \mathcal{F}_{xy}^{\alpha}(\bm{k})\\
    &&\times \left( \sum_p \Theta(\mu-\varepsilon_{\alpha\bm{k}}^{(p)})-\sum_s \Theta(\mu-\chi_{\alpha\bm{k}}^{(s)}) \right),
    \label{N3_ansatz}
\end{eqnarray}
where $\mathcal{F}_{xy}^{\alpha}(\bm{k})=i(\langle \partial_{k_x}u_{\alpha\bm{k}}|\partial_{k_y}u_{\alpha\bm{k}}\rangle-\langle \partial_{k_y}u_{\alpha\bm{k}}|\partial_{k_x}u_{\alpha\bm{k}}\rangle)$ is the Berry curvature of the $\alpha$-th magnetic Bloch band, and $\varepsilon_{n\bm{k}}^{(p)}$ and $\chi_{n\bm{k}}^{(s)}$ denote the poles and zeros of the corresponding scalar Green's function. Interestingly, Eq.~\eqref{N3_ansatz} mirrors the spirit of Ref.~\cite{Neupert2012}, which proposed that the Hall conductivity of an interacting system might be captured by integrating the single-particle Berry curvature weighted by suitably defined occupation factors. Subsequent work~\cite{Simon2014,Neupert2014} showed that such one-body expressions fail to reproduce the true many-body Hall response in generic interacting topological phases. This is fully consistent with $N_3[\mathrm{G}]$  reflecting the topology of the single-particle Green's function rather than the many-body Chern number.   

If the number of poles and zeros lying below the chemical potential for each band $\alpha$ is independent of crystal momentum, as in Fig.~\ref{fig:fig3}\textbf{(b)}, Eq.~\eqref{N3_ansatz} further reduces to

\begin{equation}
    N_3[\mathrm{G}] = \frac{A_{\mathrm{cell}}}{A}\sum_{\alpha}N_{1}[\mathrm{G}_{\alpha}] C_{\alpha},
    \label{N3_simp}
\end{equation}
where $A_{\mathrm{cell}}$ is the area of the magnetic unit cell, $A$ is the total area of the system, $C_{\alpha}$ is the Chern number of the $\alpha$-th magnetic Bloch band, and
\begin{equation}
\label{N1_Galpha}
    N_1[\mathrm{G}_{\alpha}]= A\int\frac{d^2k}{(2\pi)^2}\left( \sum_p \Theta(\mu-\varepsilon_{\alpha\bm{k}}^{(p)})-\sum_s \Theta(\mu-\chi_{\alpha\bm{k}}^{(s)}) \right)
\end{equation}
counts the net number of occupied poles minus zeros associated with band $\alpha$.
Equation~\eqref{N3_simp} is fully consistent with the numerical results obtained in the previous section from evaluating $N_3[\mathrm{G}]$ via the magnetic-field derivative of the bulk Luttinger count in a box geometry. In particular, when partially filling the lowest Hofstadter band (with Chern number $C_0=1$), we found $N_1[\mathrm{G}_0]/A \approx0.2$ at the chosen chemical potential. Since $A_{\mathrm{cell}}=5$, Eq.~\eqref{N3_simp} yields the prediction $N_{3}\approx 1$, in excellent agreement with our numerical evaluation.

\begin{figure*}[t!]
    \centering
    \includegraphics[width=0.9\linewidth]{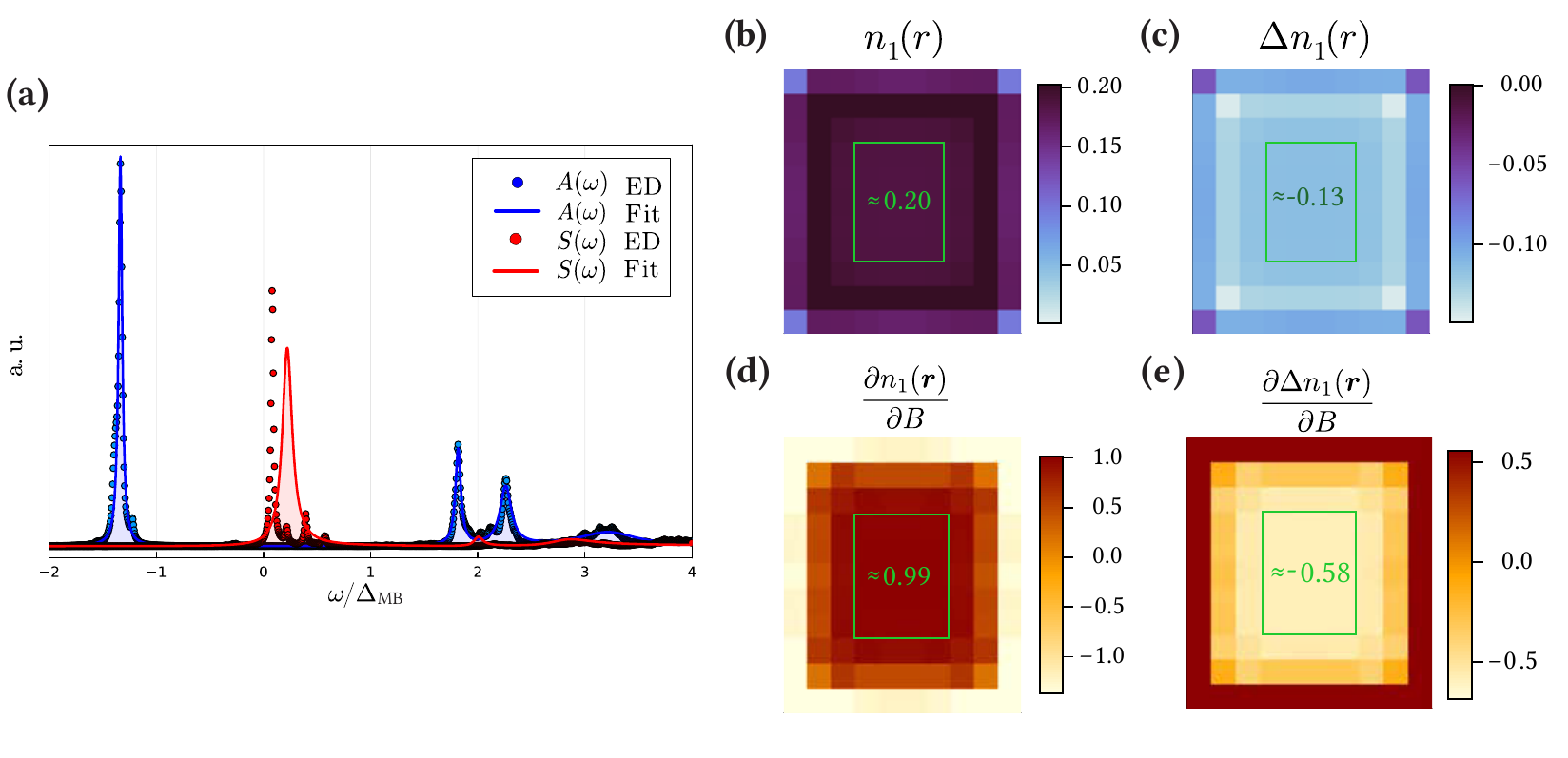}
    \caption{Extraction of the self-energy, the Luttinger count and the Luttinger integral from the spectral functions obtained with ED. \textbf{(a)} The spectral densities $A$ and $S$ defined in \cref{eq:spectral} as obtained from the exact diagonalization and extracted from the model \cref{eq:fit_spec}. Note, that $A(\omega)$ and $S(\omega)$ are plotted in different scales. \textbf{(b)} and \textbf{(c)} The distribution of the localized versions of the Luttinger count $n_1$ and the Luttinger integral $\Delta n_1$ over the lattice, as defined in \cref{eq:localn1n2}. \textbf{(c)} The St\v{r}eda response of the Luttinger count $n_1$ and the Luttinger integral $\Delta n_1$ densities. The bulk value of the approximate $\frac{\partial n_1}{\partial B}$ allows us to estimate the Ishikawa-Matsuyama invariant by $N_3 = 0.99\pm0.01$ in agreement with our numerical \cref{fig:fig4} and analytical results \cref{sec:N3 analytics}. \textbf{(d)} The fractional valued St\v{r}eda response of the Luttinger integral $\Delta N_3 = -0.58\pm0.02$ as before balances the difference between the many-body Chern number and $N_3$.}
    \label{fig:fig5}
\end{figure*}

Moreover, the results in Eqs.~\eqref{N3_simp} and~\eqref{N1_Galpha} predict a change in the value of $N_3[\mathrm{G}]$  as the chemical potential shifts to the right with respect to the in-gap Green's functions zero shown in Fig.~\ref{fig:fig3}. In that case, the Luttinger count is expected to vanish, since the contribution from each pole below the Fermi level is canceled by a zero. This change in the Luttinger count leads to the vanishing of the Ishikawa-Matsuyama invariant $N_3[\mathrm{G}]\!=\!0$ in that case, according to Eq.~\eqref{N3_simp}.

Before concluding this section, we note that Eq.~\eqref{N3_simp} becomes even simpler in the continuum Landau-level limit. In this case, the interacting Green's function within a fixed Landau level is independent of crystal momentum, $g_{\alpha\bm{k}}(z)\equiv g_{\alpha}(z)$, so all poles and zeros associated with that Landau level are dispersionless. Each Landau level carries Chern number $C_{\alpha}=1$, and the magnetic unit cell has area $A_{\mathrm{cell}}=\Phi_0/B$. Since the total flux is $\Phi = BA$, Eq.~\eqref{N3_simp} reduces to

\begin{equation}
N_3[\mathrm{G}] = \frac{\Phi_0}{\Phi}N_1[\mathrm{G}].
\label{N3_LLL}
\end{equation}
Introducing the Luttinger-count density $n_1 = N_1[\mathrm{G}]/A$ and considering that $\partial N_3[\mathrm{G}]/\partial B = 0$, we obtain from Eq.~\eqref{N3_LLL} the relation 
\begin{equation}
N_3[\mathrm{G}] = \Phi_0\frac{\partial n_1}{\partial B}\Bigg\rvert_{\mu},
\end{equation}
consistent with Ref.~\cite{peralta2023connecting}.

\section{Extracting $N_3[\mathrm{G}]$ and $\Delta N_3$ from the Local Density of States}\label{sec:scheme}

Let us discuss how our predictions can be tested experimentally. In this section we propose a scheme for extracting the Luttinger count, integral and their St\v{r}eda responses directly from the LDOS probes, which were obtained recently for FCI states \cite{farahi2023broken,hu2025high}. The values of $N_3[\mathrm{G}]$ and $\Delta N_3[\mathrm{G}]$ are the functionals of the Green's functions, self-energy and their derivatives with respect to magnetic field. In general, extracting the Green's functions and the self-energy from the local density of states is a challenging, if not impossible task \cite{valla1999evidence,jang2017full,shastry2024method}. For a Laughlin-type FCI state the scheme significantly simplifies as will be clarified below. We outline the protocol, and then, as a proof of concept, we extract $N_3[\mathrm{G}]$ and $\Delta N_3[\mathrm{G}]$ from the local density of states that were obtained numerically.

The scheme exploits several characteristic features of the Laughlin-type FCI's Green's functions. In the continuum LL limit, the spatial and  temporal dependencies of the Green's functions are disentangled due to their momentum independence. This limit is well realized for the parameters used in~\cref{fig:fig3}. In this regime, the frequency dependence of the Green's functions is fully determined by the local density of states, while the spatial dependence of the Green's functions is governed solely by the non-interacting single-particle density matrix. Consequently, the Green's functions projected onto the LHB assume the form of \cref{ansatz}, with a momentum independent function $g$:
\begin{equation}
\hat{\mathrm{G}}_{\bm{k}}(z) =  g(z)\,
|u_{\bm{k}}\rangle\langle u_{\bm{k}}|.
\label{eq:Laughlin_ansatz}
\end{equation}
Here we dropped the band index $\alpha$, neglecting the mixing with the higher Hofstadter bands. The spectral functions provide us with the imaginary part of the function $g(\omega)$. The real part can be obtained from the Kramers-Kronig relations. Thus, the Green's functions $\mathrm{G}(r,r;z)$ can be obtained, provided that the non-interacting Bloch vectors $|u_{\bm{k}}\rangle$ are known.

An additional simplification stems from the observation that the spectral function of the Laughlin-type FCI's is composed of a small number of sharp coherent peaks well separated from the incoherent background; see~\cref{fig:fig3} and Refs~\cite{rezayi1987electron,Jain2005}. This motivates a simple model for the spectral functions consisting of a sum of a few narrow Lorentzians, representing the coherent part of the spectrum and an additional broader Lorentzian peak representing the incoherent background:
\begin{align}
A_{\mathrm{fit}}(\omega) &=  A_{\mathrm{coh}}(\omega)+A_{\mathrm{inc}}(\omega),\\
A_{\mathrm{coh}}(\omega) &= \sum_{k=1}^{N_p} Z_k \frac{\gamma_k^2}{(\omega-\omega_k)^2+\gamma_k^2},\\
A_{\mathrm{inc}}(\omega) &=\, Z_b \frac{\Gamma^2}{(\omega-\Omega)^2+\Gamma^2}.
\label{eq:fit_spec}
\end{align}
Here $N_p$ is the number of coherent peaks, $Z_k$ are the spectral weights of the coherent peaks, $\omega_k$ and $\gamma_k$ are their positions and widths; $\Omega$ and $\Gamma$ represent the position and width of the incoherent peaks, respectively. Fitting the spectral functions with $A_{\mathrm{fit}}(\omega)$, one readily obtains the approximate frequency dependent part of the Green's functions:
\be
g_{\mathrm{fit}}(\omega) = \sum_{k=1}^{N_p} \frac{\pi Z_k \gamma_k}{\omega - \omega_k + i \gamma_k} + \frac{\pi Z_b \Gamma}{\omega - \Omega + i \Gamma}.
\label{eq:gffit}
\ee

We now apply the procedure to the spectral functions that were numerically obtained for the $\nu\!=\!\frac{1}{3}$ FCI state; see~\cref{fig:fig3}. For this state we observed the three dominant coherent peaks $N_p = 3$, one corresponding to the hole-like state, and the other two corresponding to the particle-like side of the spectrum. We fitted the data with the model spectral functions in \cref{eq:fit_spec}. Since we are ultimately interested in the response of the particle density to a small perturbing magnetic field, we included in the minimization procedure a Lagrange multiplier penalizing the deviation of the area under the peaks from the area under the data peaks. Once the approximate $g_{\mathrm{fit}}(\omega)$ are obtained, the Green's functions can be calculated from \cref{eq:Laughlin_ansatz}. Thereafter, we evaluate numerically the derivatives of the Green's functions and the self-energy with respect to magnetic field, repeating the protocol for the spectral functions of the same system in a small perturbing magnetic field. Finally we evaluate the local densities of the Luttinger count, the Luttinger integral  \cref{eq:localn1n2} and the corresponding St\v{r}eda responses.

The resulting approximate spectral functions $A_{\mathrm{fit}}(\omega)$ and $S_{\mathrm{fit}}(\omega)$ \cref{eq:spectral} along with the distributions of the Luttinger count $n_1(\bm{r})$, the Luttinger integral $\Delta{n_1(r)}$ and their derivatives with respect to the magnetic field are presented in \cref{fig:fig5}. The approximate and the exact results presented in Figs.~\ref{fig:fig5} and \ref{fig:fig4}
respectively are in a remarkable agreement, given the simplicity of the model Green's functions. The self-energy peak $S_{\mathrm{fit}}(\omega)$ in the many-body gap presented in panel \textbf{(a)} of \cref{fig:fig5} deviates only slightly from the exact result. The situation can be further improved by a more adequate model for the incoherent part of the spectral function $A_{\mathrm{inc}}$. We found that, while the position of the peak is quite robust to the fine details of the model for $A_{\mathrm{inc}}$, the position does depend on the spectral weight $Z_b$ of the background. The bulk values of the approximate Luttinger count $n_1 = 0.2000\pm0.0001$, and the Luttinger integral $\delta n_1=0.131\pm0.002$ are in a good agreement with the ED results \cref{fig:fig4} and the theoretical predictions.

Most importantly, the protocol allowed us to estimate the values of the Ishikawa-Matsuyama invariant $N_3[\mathrm{G}]$  and the correction $\Delta N_3[\mathrm{G}]$ from the local density of states, which can be directly accessed experimentally. For the same choice of the chemical potential as in \cref{sec:Results_3} we obtain $N_3 = 0.99\pm0.01$ and $\Delta N_3 = -0.58\pm0.02$ in agreement with the ED results, whereas for the complex integration contour enclosing the bulk zero $N_3 = 0.001\pm0.009$ and $\Delta N_3 = -0.39\pm0.02$. Let us note that the experimental data for the spectral functions $A(\omega)$ reported in \cite{farahi2023broken,hu2025high} exhibit a more complicated structure than the ED results used in our tests. A convolution with the DOS of the tip obscures the sharp coherent picture among other factors inherent to a real experiment. Nevertheless, this proof-of-concept implementation of the protocol gives us a ground for a cautious optimism regarding the experimental assessment of our results.

\section{Concluding Remarks}
\label{sec:conclusions}

In this work we investigated single-particle Green's functions and their associated topological invariants, the Luttinger count and Ishikawa-Matsuyama invariant, for a Laughlin-type FCI state. By explicitly evaluating the Luttinger count and the Luttinger integral \cite{luttinger1960ground}, we demonstrated the breakdown of  Luttinger's theorem. Using the connection between Green's function invariants and St\v{r}eda responses~\cite{peralta2023connecting}, we evaluated $N_3[\mathrm{G}]$ and its deviation from the many-body Chern number, $\Delta N_3[\mathrm{G}]$, as the magnetic responses of the Luttinger count and the Luttinger integral, respectively. We found that the fractional nature of the St\v{r}eda response, characteristic of FCI states, is entirely captured by the response of the Luttinger integral, while $N_3[\mathrm{G}]$ itself encodes only the single-particle topology inherited from the underlying magnetic Bloch bands and the momentum resolved values of the Luttinger count.
Finally, we proposed a scheme directly relating the experimentally accessible local density of states \cite{farahi2023broken,hu2025high} to the Luttinger count, integral and their corresponding St\v{r}eda responses $N_3[\mathrm{G}]$ and $\Delta N_3[\mathrm{G}]$.

Our results raise several intriguing directions for future work. One appealing scenario is an interface between two FCI states with different quantized values of the Ishikawa-Matsuyama invariant. According to the bulk-boundary correspondence of Ref.~\cite{gurarie2011single}, the mismatch in $N_3$ should be reflected in the Luttinger count of the boundary Green's function, so that dispersive branches of zeros or poles of the boundary Green's functions are expected to emerge and cross the chemical potential. An especially interesting case occurs when two FCI states share the same many-body Chern number but differ in their Ishikawa-Matsuyama invariant due to a shift in the chemical potential. Even if the boundary remains gapped, the analytical structure of the boundary Green's functions is expected to undergo qualitative changes~\cite{gurarie2011single}, pointing toward interfacial phenomena beyond those captured by conventional many-body topology~\cite{wagner2023mott}.

Addressing these questions may require complementary numerical and analytical methods. Tensor-network methods~\cite{white1992density, jordan2008classical}, which have already proven effective for FCI studies \cite{feiguin2008density,Grushin2015,Motruk2015,He2017,Motruk2017,
Dong2018,Rosson2019,Schoonderwoerd2019,
Palm2021,Andrews2021,Boesl2022,Weerda2024,van2025fractional}, could be adapted with controlled accuracy~\cite{long2024spectra} through benchmarking against exact diagonalization. Diagrammatic Monte-Carlo methods~\cite{prokof1998polaron}
have recently revealed signatures of FQH physics in the Green's functions~\cite{currie2024fractional}, while analytical approaches based on composite-fermion theories~\cite{yue2024electronic} or the thin torus limit~\cite{adhidewata2025excitation} may further elucidate the single-electron structure of fractionalized topological matter.  \\     

\paragraph*{Acknowledgments}
The authors warmly thank Antoine Georges for his insightful feedback on the manuscript and for discussions. They also acknowledge discussions with Victor Gurarie, Matteo Rizzi, Laurens Vanderstraeten and Botao Wang. This research was partly financially supported by the ERC Grant LATIS, the FRS-FNRS (Belgium), the EOS project CHEQS, and the Fondation ULB. A.A.M and A.M.N also acknowledge the financial support from the Russian Quantum Center in the framework of the Russian Quantum Technologies Roadmap. L.P.G. acknowledges support provided by the FRS-FNRS Belgium and the L’Or\'eal-UNESCO for Women in Science Programme.

\paragraph*{Contribution statement} A.A.M. conceived the project with inputs from L.P.G., N.G and A.M.N. The numerical simulations were performed and lead by A.M.N, with contributions from A.A.M. The analytical calculations were performed by L.P.G. and N.R.C. All authors analyzed and discussed the results. A.A.M., L.P.G., N.G. and A.M.N. wrote the manuscript with inputs from N.R.C. The research was supervised by L.P.G., N.G. and A.A.M.

\appendix
\section{Numerical Evaluation of the Relevant Quantities}
\label{sec:ap_numerics}

\subsection{Green's functions and self-energy}

In this appendix we present details of the numerical evaluation of the Green's functions for the FCI state. The numerical calculation of the derivatives of the Luttinger count and the Luttinger integral \cref{eq:localn1n2} with respect to the magnetic field requires working with open boundary conditions. This is a consequence of the Dirac quantization of the magnetic field under periodic boundary conditions, which does not allow for infinitesimally small variations of the magnetic field for a finite system. The extraction of bulk Green's functions under open boundary conditions is possible only when the system size is large compared to the correlation length. To make the calculations feasible, we therefore consider the Harper–Hofstadter–Hubbard model projected onto the lowest Hofstadter band \cref{eq:Hamiltonian}.

We start by diagonalizing the non-interacting Hofstadter Hamiltonian for a given geometry:
\be
\label{eq:ap_HH_ham}
\begin{split}
H_0 &= \sum_{n,m} 
  -t \left( 
    c^{\dagger}_{n,m+1} c_{n,m} 
    + c^{\dagger}_{n+1,m} c_{n,m} e^{i 2\pi m \varphi} 
    + h.c. 
  \right)\\
  &\equiv \sum_{ij}h_{ij}c^\dagger_ic_j=\sum_{\beta=1}^{N_s} \ve_{\beta} c^\dagger_\beta c_\beta,
\end{split}
\ee
where $N_s$ is the number of sites and the operators $c_j$ and $c_\beta$ are related by $c_j = \sum_\beta U_{j,\beta}c_\beta$ with $U_{j,\beta}$ being the matrix of eigenvectors of the single-particle Hamiltonian $h_{ij}$. The nearest-neighbor Hubbard interaction
\be
\label{eq:ap_Hubbard}
H_{int} = \sum_{<ij>}V \;: n_i n_j:
\ee
 can then be expressed in terms of the operators $c_\beta$. The normally ordered projected Hamiltonian in \cref{eq:Hamiltonian} is obtained by truncating the sums over $\beta$, keeping only the operators corresponding to the lowest Hofstadter band. The resulting Hamiltonian in the Fock space generated by the operators $c_\beta$ is then constructed and diagonalized numerically using the Lanczos algorithm \cite{lanczos1950iteration}. In our implementation, we employ the Arpack.jl Julia package, which provides a Julia wrapper around the Fortran ARPACK library.
 
The Green's functions are evaluated numerically using the Lehmann representation \cite{altland2010condensed}, which at zero temperature reads
\be
\label{eq:ap_lehmann}
\begin{split}
G_{\gamma, \beta}(z) \smeq \sum_m\frac{\la 0| c_\gamma | m^{N+1}\ra \la m^{N+1}| c^\dagger_\beta | 0\ra }{z - (E^{N+1}_m - E_0) }\\+ \sum_m\frac{\la 0| c^\dagger_\beta | m^{N-1}\ra \la m^{N-1}| c_\gamma | 0\ra }{z + (E^{N-1}_m - E_0)}. 
\end{split}
\ee
Here $|m^{N\pm1}\ra$ and $E^{N\pm1}_m$ denote the eigenstates and eigenvalues of the Hamiltonian \cref{eq:Hamiltonian} for $N\pm1$ particles. The procedure starts by obtaining the ground-state vector $|0\ra$ for a fixed number of particles $N$, after which the states $c^{(\dagger)}_\beta|0\ra$ are constructed. The key step in the Lanczos-based numerical evaluation of the Green's functions is the approximation of the resolvent operator $\frac{1}{z\mathbb{1} - H}$ by its projection onto the Krylov space $\frac{1}{z\mathbb{1} - \check{H}}$~\cite{koch20118}. If the Krylov space $\mathcal{K}$ has dimensionality  $M_{\mathcal{K}}$, the Krylov projected operator $\frac{1}{z\mathbb{1} - \check{H}}$ reproduces the same first $2M_{\mathcal{K}}+1$ moments $\int d\omega G_{\beta\beta}(\omega)\omega^m$ of the exact operator $\frac{1}{z\mathbb{1} - \check{H}}$~\cite{koch20118}. Therefore, although the Lanczos algorithm does not provide accurate approximations for the highly excited states $|m^{N\pm1}\ra$, the Green's functions obtained in this way converge rapidly to the exact ones as $M_{\mathcal{K}}$ increases.

Once the Green's functions are obtained in the eigenbasis of the Harper-Hofstadter Hamiltonian, the self-energy is computed using the Dyson equation: $G(z)^{-1} = G^{-1}_0(z)-\Sigma(z)$. Importantly, the order of matrix-inversion and the transformation to the real-space basis matters. The self-energy must first be evaluated in the eigenbasis of the Harper-Hofstadter Hamiltonian and only afterwards transformed to the real-space basis. This is due to the mismatch between the sizes of the two bases arising from the projection onto the lowest Hofstadter band. In particular, the Green's function expressed in the real-space basis has rank equal only to the number of states in the LHB, making its inversion ill-defined.

\begin{figure}[t]
    \centering
    \includegraphics[width=0.9\columnwidth]{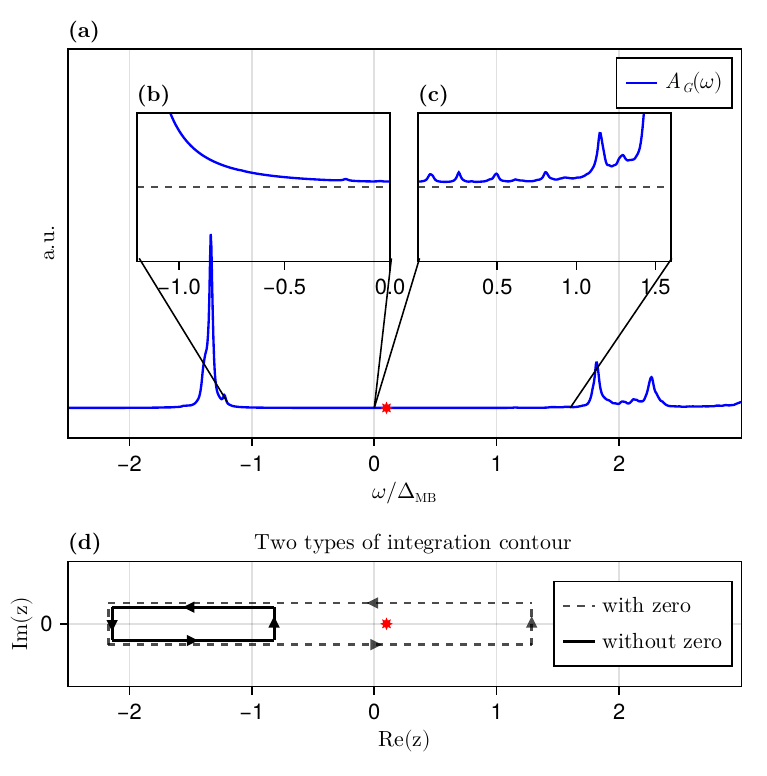}
    \caption{\textbf{(a)} Local density of states in the bulk. Insets \textbf{(b,c)} show the density of states on a smaller scale below and above $\omega=0$, respectively. \textbf{(d)} Schematic illustration of the possible integration contours for evaluating the local Luttinger count and the Luttinger integral, \cref{eq:localn1n2}. The red star indicates the position of the in-gap bulk Green's function zero. Because edge-induced poles of the Green's function appear above $\omega=0$, the numerical integration is performed only along the contour shown in solid line.}
    \label{app:fig:integration_contours}
\end{figure}

\subsection{Evaluation of $n_1(r)$ and $\Delta n_1(r)$.}
\label{sec:ap_countour}

We next compute the localized versions of the Luttinger count and the Luttinger integral, defined in \cref{eq:localn1n2}, and evaluate their numerical derivative with respect to magnetic field. Here we summarize the procedure used in the calculations.

The integration contour over continuous Matsubara frequencies appearing in the definitions of the Luttinger count and the Luttinger integral (\cref{eq:n1_n2}) can be closed in the lower half of the complex-frequency plane. The contour can then be deformed into the shape presented schematically in \cref{app:fig:integration_contours} \textbf{(d)}. During this deformation, no Green's function pole or zero should cross the contour, as doing so would make the integrand in \cref{eq:localn1n2} singular and prevent convergence. As discussed in the main text, the chemical potential can be chosen arbitrarily within the charge gap. Depending on this choice, the single zero of the bulk Green's function lies either within or outside the integration contour. In practice, stable numerical convergence is achieved only for contours that do not enclose the zero. The convergence crucially depends on the regularity of the Green's functions and self-energies close to the points where the contour crosses the real-frequency axis. As demonstrated in \cref{app:fig:integration_contours} \textbf{(a-c)} the spectral density develops additional peaks inside the charge gap to the right of the bulk Green's function zero. These additional peaks are absent in the spectral densities in periodic boundary conditions and can be attributed to in-gap edge contributions to the density of states, as detailed in Appendix~\ref{sec:app:finite_size}.

\section{Finite-Size Effects}
\label{sec:app:finite_size}

Throughout the work we aim to characterize the bulk properties of the FQH droplet. The finite size of our system affects the results in two distinct ways. First, under open boundary conditions, edge and bulk physics may mix due to the finite spatial extent of the system. Second, the limited size can modify the bulk properties themselves. The latter effect is present even in calculations performed with periodic boundary conditions. To quantify both types of finite-size effects, we begin by analyzing the decay of correlation functions.

\begin{figure}
    \centering
    \includegraphics[width=0.9\columnwidth]{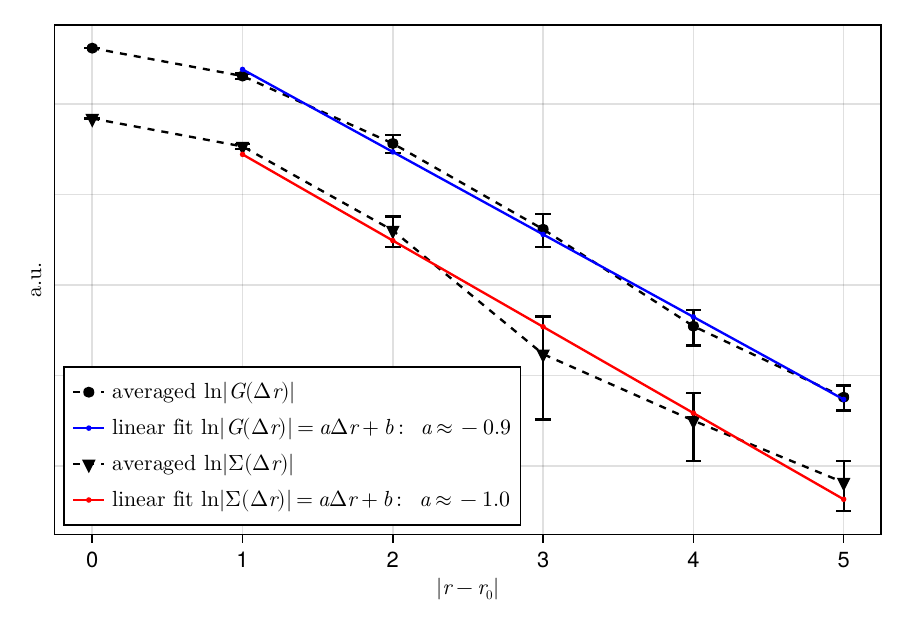}
    \caption{\textbf{Spatial decay of the Green's function and self-energy.} Dashed curve with circle (triangle) markers shows the Green's function (self-energy) behavior averaged over different paths from the bulk region toward the edge. The data are shown on a logarithmic $y$-axis. Solid lines indicate linear fits yielding a correlation length of $l_{\text{cor}}\sim 1$ lattice constant.}
    \label{fig:gf_decay}
\end{figure}

Correlation functions in the bulk of the FCI state decay at least exponentially fast\footnote{In fact, for a Laughlin-type FCI state one may expect correlations to decay faster than exponentially.}, since all bulk excitations are gapped. Figure~\ref{fig:gf_decay} shows the spatial decay of the Green's functions and self-energies at zero frequency on a logarithmic scale. The linear fits (solid lines) demonstrate that the Green's functions can be bounded by $\mathrm{G}(\bm{r},\bm{r}';\omega) < e^{-|\bm{r}-\bm{r}'|/l_{\mathrm{cor}}}$ with a correlation length of the order of one lattice constant, $l_{\mathrm{cor}} \approx 1$. That this bound applies at arbitrary frequency follows from the decomposition in \cref{eq:Laughlin_ansatz}, which states that the temporal and spatial dependencies of the Green's functions factorize with a good accuracy for an almost flat LHB. Another immediate consequence of the decomposition in \cref{eq:Laughlin_ansatz} is that the decay of Green's functions is determined by the noninteracting Hamiltonian. The obtained correlation length $l_{\mathrm{cor}} \approx 1$ is compatible with the magnetic length at the flux $\varphi = \frac{1}{5}$: $l_B = \sqrt{\frac{q}{2\pi}}\approx 0.9$.
\begin{figure}
    \centering
    \includegraphics[width=0.95\columnwidth]{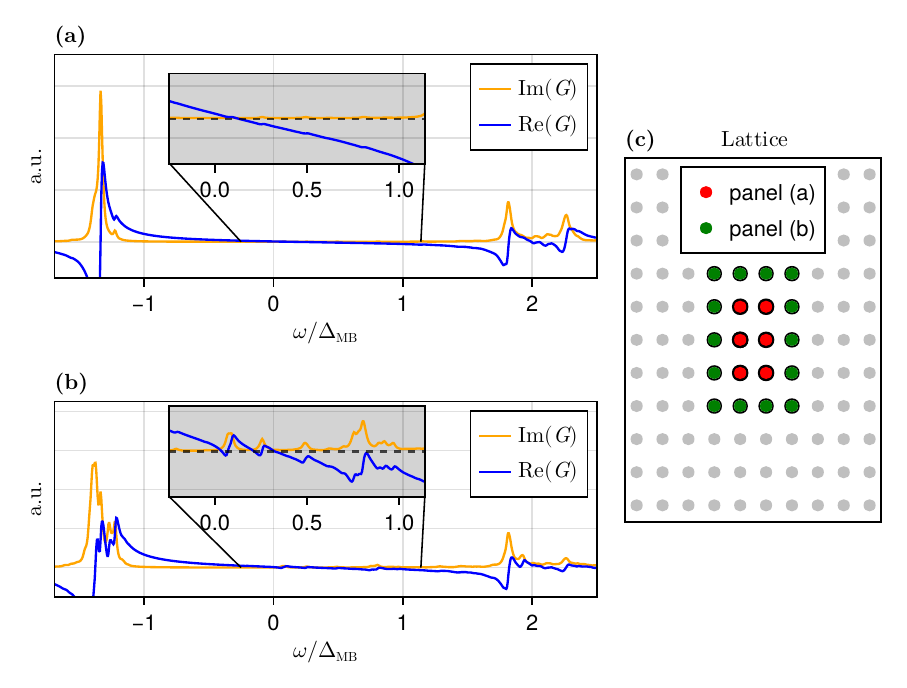}
    \caption{Illustration of the influence of edge-induced poles and zeros of the Green's function on the numerical evaluation of $n_1$ and $\Delta n_1$. \textbf{(a)} Real and imaginary parts of the diagonal elements of the Green's functions in the center of the sample, highlighted in red in panel \textbf{(c)}. \textbf{(b)} Real and imaginary parts of the diagonal Green's function averaged over the region highlighted in green in panel \textbf{(c)}, where multiple poles and zeros appear.}
    \label{fig:gf_decay_in_detail}
\end{figure}

Let us first discuss the finite-size effects arising from the influence of the metallic edges. As anticipated in the previous section, in-gap poles of the Green's functions associated with the edges affect the convergence of the numerical evaluation of $n_1$, $\Delta n_1$, and their St\v{r}eda responses. The convergence of the integration relies on the regularity of the integrand in \cref{eq:localn1n2}. In particular, no poles or zeros of the Green's functions matrix elements $\mathrm{G}(r,r';\omega)$ should appear in the vicinity of the points where the integration contour intersects the real-frequency axis. As demonstrated in \cref{fig:gf_decay_in_detail}, this condition is not satisfied for contours enclosing the single bulk Green's function zero.

This zero, located within the single-particle gap, is visible in Fig. \ref{fig:gf_decay_in_detail}(\textbf{a}) as the point where both the real and imaginary parts of the Green's function $\mathrm{G}(\bm{r}_0,\bm{r}_0;\omega)$ vanish for any site $\bm{r}_0$ within the central red region in Fig.~\ref{fig:gf_decay_in_detail} (\textbf{c}). However, moving just one lattice site toward the edge (green region in Fig. \ref{fig:gf_decay_in_detail}(\textbf{c})) and averaging over this larger region already reveals additional poles and zeros. Despite the short correlation length of order one lattice constant these additional divergencies 
spoil the convergence of the numerical integration for contours enclosing the bulk zero. In practice, we therefore restrict the integration to contours that cross the real axis below the bulk zero, ensuring stable evaluation of the Luttinger count and the Luttinger integral.

\begin{figure*}
    \centering
    \includegraphics[width=0.9\linewidth]{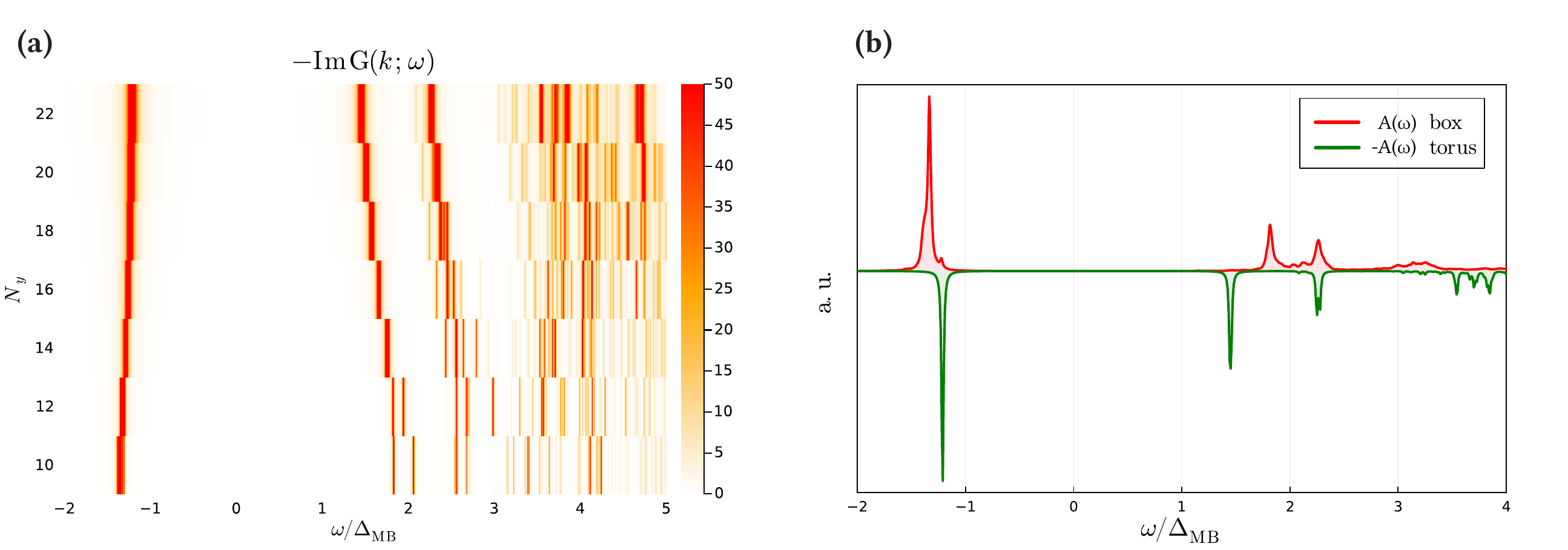}
    \caption{\textbf{(a)} Spectral densities $A(\omega)$ for different system sizes, obtained by increasing $N_y$ while keeping $N_x=9$ constant. The number of particles is fixed to $N = 6 $, while the magnetic flux per plaquette is $\varphi=1/q$ with $q=N_y/2$, ensuring a constant filling fraction of $\nu=1/3$. \textbf{(b)} Comparison between the spectral densities $A(\omega)$ for a torus of size $N_y=22$, $N_x=9$ and for a box with $N_y=11, N_x = 10$.}
    \label{fig:app:box vs torus}
\end{figure*}

\begin{figure}[b]
    \centering
    \includegraphics[width=0.9\columnwidth]{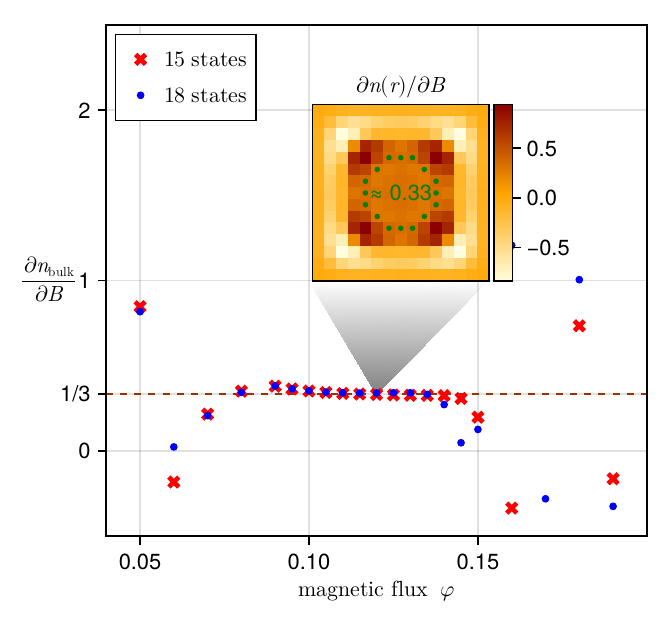}
    \caption{Stabilization of the St\v{r}eda response in a larger system with $15\times 15$ sites and 6 particles. The many-body Chern number extracted from the St\v{r}eda response as a function of the magnetic flux per unit cell exhibits a clear plateau corresponding to the $1/3$ FCI in the range $0.08<\varphi<0.14$ (in units of the flux quantum $\Phi_0$). The inset shows the spatially resolved density response, yielding a bulk Chern number of $C_{\text{MB}}=0.33 \pm 0.01$. Red crosses and blue circles correspond to calculations with different number of bulk states retained in the LHB projection.}
    \label{fig:streda_response_bigger_system}
\end{figure}

The second type of finite-size arises from the limited system size itself, which modifies the bulk properties even in the absence of edges, as realized on the torus.
In particular, the spectral densities shown in \cref{fig:fig3} \textbf{(a)} obtained with periodic (torus) and open (box) boundary conditions are visibly different. The double-peak structure on the particle side of the spectral functions, predicted in Ref.~\cite{Jain2005}, is evident in the box geometry, whereas on the torus each peak appears to be split into several smaller peaks. We argue that the general structure of the spectral functions on torus becomes compatible with the one obtained in the bulk of the box geometry as the system size increases. To support this, we computed the spectral densities on the torus for larger systems with a fixed particle number $N = 6$. The resulting densities of states are presented in \cref{fig:app:box vs torus}, where the characteristic double-peak structure followed by an incoherent background emerges clearly for larger system sizes, matching the bulk behavior seen in the box geometry.

Let us close the section by discussing finite-size effects on the St\v{r}eda response in the FQH state. Figure~\ref{fig:fig4} panel \textbf{(e)} of the main text shows the onset of the plateau in the St\v{r}eda response $\frac{\partial n}{\partial B}$ as a function of the magnetic flux per unit cell, near the chosen value $\varphi = 1/5$. We verified that the plateau stabilizes in a larger system of $15\times15$ sites, as shown in \cref{fig:streda_response_bigger_system}.
Since the exact number of bulk states in the LHB cannot be determined precisely in the box geometry, we computed the response using both 15 and 18 states in the LHB and observe that a clear plateau emerges in both cases.

\section{Momentum Dependence of Ground State Correlators on a Torus}
\label{sec:ap single gs}
\begin{figure*}
    \centering
    \includegraphics[width=0.9\linewidth]{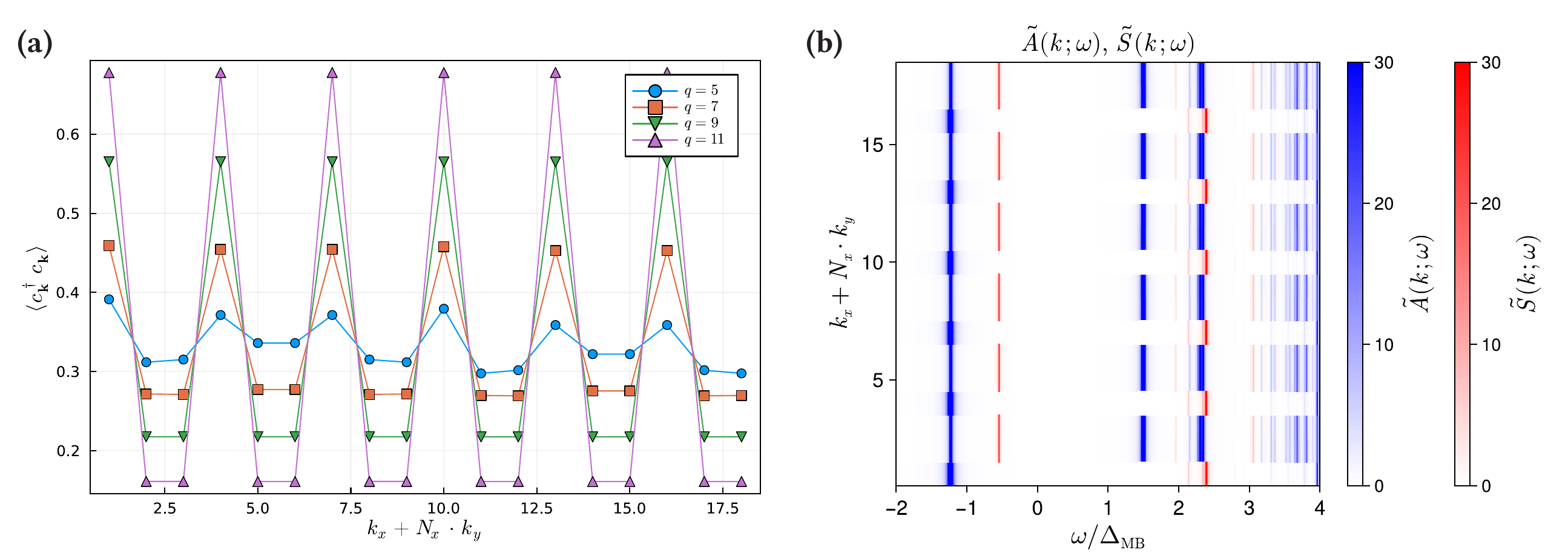}
    \caption{\textbf{(a)} Momentum-state occupations $\langle\Psi^1_0|c^\dagger_k c_k^{}|\Psi^1_0\rangle$ for one of the degenerate ground states $|\Psi^1_0\rangle$ on the torus for different values of the magnetic flux per unit cell $\varphi = 1/q$. The torus has dimensions $N_y = 2q$ and fixed $N_x=9$. Increasing $q$ increases the magnetic length $\ell_B$, bringing the system closer to the thin-torus limit. \textbf{(b)} Spectral functions of the Green's function, $\tilde{A}(\omega,\bm{k})$, and of the self-energy, $\tilde{S}(\omega,\bm{k})$, defined in \cref{eq:app:signgle_gs_GF}, shown for $q=11$.}
    \label{app:fig:thin torus}
\end{figure*}

As demonstrated in the inset of \cref{fig:fig3} \textbf{(b)}, the spectral functions $\tilde{A}(k,\omega)$ calculated by choosing a single ground state:
\be
\label{eq:app:signgle_gs_GF}
\begin{split}
 \tilde{A}(k,\omega) = \sum_n  \bigl[ & |\langle \Psi_0^1|\hat{c}_k|\Psi_n\rangle|^2\delta(\omega -E_n + E_0) + \bigr.\\
    \bigl. & |\langle \Psi_0^1|\hat{c}^\dagger_k|\Psi_n\rangle|^2\delta(\omega + E_n - E_0) \bigr],
\end{split}
\ee
feature a momentum dependence that is absent for the thermodynamically averaged spectral functions $A(k,\omega)$ \cref{eq:spectral}. There is a clear periodic structure in the positions of the Green's functions zeros and poles with a period $2\pi/N_x\cdot3$. In this section we argue that this peculiar momentum dependence arises naturally in the thin torus limit~\cite{tao1983fractional,seidel2005incompressible,bergholtz2005half, bergholtz2008quantum} and persists — albeit gradually disappearing — as the torus circumference increases. 

FQH states in the continuum can be adiabatically connected to Tao-Thouless crystal states~\cite{tao1983fractional,seidel2005incompressible,bergholtz2005half,bergholtz2008quantum,adhidewata2025excitation}. Remarkably, these crystalline states are exact solutions of the problem of a partially filled Landau level in the thin-torus or cylinder limit. Let us outline the main steps of the procedure. Consider a $2d$ electronic system in a uniform magnetic field $B$ in the Landau gauge $\mathbf{A} = - (By, 0)$ on a torus. We assume that the torus is infinite in the $y$-direction $L_y \to \infty$ and finite in the explicitly translationally invariant $x$-direction. The single-electron wave functions in the lowest Landau level are plane waves in the $x$-direction and Gaussian localized in the $y$-direction:
\be
\label{app:eq:landau wf}
\psi_k(x,y) \sim e^{i k x} \exp\!\left[-\frac{1}{2\ell_B^2} \left(y - k \ell_B^2\right)^2 \right].
\ee
Here $\ell_B = \sqrt{\frac{\hbar c}{e B}}$ is the magnetic length. The centers of localization of the states $\psi_k(x,y)$ are equally spaced in $y$-direction with a spacing $2 \pi \ell_B^2/L_x$.

A general density-density interaction projected onto the lowest Landau level takes the form\cite{bergholtz2008quantum}:
\be
\begin{split}
\label{app:eq:interaction}
\hat{H}_{int} &= \sum_{\{k_i\}}V_{k_1,k_2,k_3,k_4}c^\dagger_{k_1}c^\dagger_{k_2} c^{}_{k_3}c^{}_{k_4},\\
V_{k1,k2,k3,k4} &= \int d^2\bm{r}_1 d^2\bm{r}_2 \psi^*_{k_1}(\bm{r_1})\psi^*_{k_2}(\bm{r_2})\times\\ &\times V(|\bm{r_1} - \bm{r_2}|)\psi_{k_3}(\bm{r_2})\psi_{k_4}(\bm{r_1}).
\end{split}
\ee
In the limit $L_x/\ell_{B}\to 0$ the overlap between the wavefunctions $\psi_{k_1}(x,y)$ and $\psi_{k_2}(x,y)$ becomes exponentially small. As a result, the only surviving terms in \cref{app:eq:interaction} correspond to simple electrostatic repulsion between the fermions on a quasi-one-dimensional lattice. Since the Landau level is flat, the Hamiltonian in this limit reduces to
\be
\label{app:eq:TT}
\hat{H}_{TT} = \sum_{i,k}V_k n_i n_{i+k}.
\ee
At a filling fraction $\nu = 1/(2m+1)$ the energy is minimized by a crystalline state in which the fermions are separated from each other as far as possible. At the filling $\nu=1/3$ the ground states in the occupation basis are: 
\be
\label{app:eq:TTGS}
|\Psi^1_0\rangle = |...001001001...\rangle.
\ee  
The three degenerate ground states are related to each other by translations of this pattern and therefore carry different total momenta. The equal-time correlators $\langle\Psi^1_0|c^\dagger_k c_k^{}|\Psi^1_0\rangle$ are therefore momentum-dependent and exhibit a $\frac{2 \pi}{L_x}\cdot 3$ periodicity. Outside the strict Tao--Thouless limit ${L_x/\ell_{B}\to0}$, the scattering between states with different $k$ cannot be neglected and the density starts to fluctuate. However, the modulation of the occupation of the momentum states $\langle\Psi^1_0|c^\dagger_k c_k|\Psi^1_0\rangle$ persists at any finite $L_x$.

The Laughlin-type FCI states exhibit essentially the same physics \cite{bernevig2012thin}. \cref{app:fig:thin torus} \textbf{(a)} shows the momentum-state occupations for a single ground state and different values of the magnetic flux per plaquette $\varphi =1/q$ in a torus with dimensions $N_y=2q$, $N_x=9$ and a fixed number of particles $N=6$. Increasing the value of $q$ increases the magnetic length and therefore brings the system closer to the thin torus limit. As a result, the modulation of the momentum-states occupation becomes more pronounced with increasing $q$ and persists up to $q=5$, corresponding to the results presented in the main text. The momentum dependence of the spectral densities $\tilde{A}(\omega,\bm{k})$ and  $\tilde{S}(\omega,\bm{k})$ becomes respectively more pronounced for larger $q$ as illustrated in \cref{app:fig:thin torus} \textbf{(b)}.

\section{Derivation of Eq.~\eqref{N3_ansatz}}
\label{sec:ap derivation N3_ansatz}

We present here additional details on the derivation of Eq.~\eqref{N3_ansatz} in the main text, which provides a simplified expression for the Ishikawa-Matsuyama invariant when the single-particle Green's function is approximated as diagonal in the Bloch band index. Substituting Eq.~\eqref{ansatz} into Eq.~\eqref{eq:N3}, and after some algebra, one obtains
\begin{align}
\notag
&N_3[\mathrm{G}]=\!-\frac{\epsilon^{0\nu\rho}}{8\pi^2}\!\!\sum_{\alpha} \!\!\int\!\!\!dz e^{z0^{+}}\!\!\!\!\int \!\!d^2 k\Bigg[\!2g_{\alpha\bm{k}}^{-1} \frac{\partial g_{\alpha\bm{k}}}{\partial z}\langle \partial_{k_{\nu}}u_{\alpha\bm{k}}|\partial_{k_\rho}u_{\alpha\bm{k}}\rangle\\
\notag
\!\!\!\! &\!-\!\sum_{\beta}\!\left(\!\frac{\partial g_{\alpha\bm{k}}}{\partial z}g_{\beta\bm{k}}^{-1} -\frac{\partial g_{\alpha\bm{k}}^{-1}}{\partial z}g_{\beta\bm{k}}\!\right)\langle \partial_{k_{\nu}} u_{\alpha\bm{k}}|u_{\beta\bm{k}}\rangle \langle u_{\beta\bm{k}}|\partial_{k_{\rho}}u_{\alpha\bm{k}}\rangle\!\Bigg],\\
\label{aux}
\end{align}
where summation over repeated indices $\nu,\rho$ is implicit. The frequency integral in the first term of Eq.~\eqref{aux} can be evaluated explicitly using the meromorphic structure of the diagonal Green's function elements, 
\begin{equation}
    g_{\alpha\bm{k}}(z) = \frac{\prod_s (z-\chi_{\alpha\bm{k}}^{(s)})}{\prod_p (z-\varepsilon_{\alpha\bm{k}}^{(p)})},
    \label{mero_g}
\end{equation}
where $\varepsilon_{\alpha\bm{k}}^{(p)}$ denotes their poles and $\chi_{\alpha\bm{k}}^{(s)}$ their zeros. Using Eq.~\eqref{mero_g}, one readily finds
\begin{eqnarray}
\notag
\!\!\int\!\!\!\!dz e^{z 0^{+}}\!\!g_{\alpha\bm{k}}^{-1}(z) \frac{\partial g_{\alpha\bm{k}}}{\partial z}
    \!&=&\!\!\!\int\!\!\!dz e^{z0^{+}} \left(\sum_s  \frac{1}{z-\chi_{\alpha\bm{k}}^{(s)}}-\sum_p\frac{1}{z-\varepsilon_{\alpha\bm{k}}^{(p)}}\right)\\
    \notag
    \!&=&\!\!2\pi i\!\left(\!\sum_s\!\Theta(\mu-\chi_{\alpha\bm{k}}^{(s)})-\sum_p\!\Theta(\mu-\varepsilon_{\alpha\bm{k}}^{(p)})\!\right),\\
    \label{aux2}
\end{eqnarray}
where we have used that the contour integration in the complex frequency plane selects the residues of poles and zeros lying below the chemical potential $\mu$. After performing an integration by parts in the last term of Eq.~\eqref{aux} and inserting Eq.~\eqref{aux2}, one arrives at
\begin{align}
\notag
    &N_3[\mathrm{G}]\!=\!\sum_\alpha\!\!\int\!\!\frac{d^2k}{2\pi}   \mathcal{F}_{xy}^{\alpha}(\bm{k})\!\left(\!\sum_p \Theta(\mu-\varepsilon_{\alpha\bm{k}}^{(p)})\!-\!\sum_s \Theta(\mu-\chi_{\alpha\bm{k}}^{(s)})\!\right)\\
    \notag
    &+\frac{\epsilon^{0\nu\rho}}{8\pi^2}\sum_{\alpha\beta}\int\!\!d^2 k \mathcal{B}_{\alpha\beta}(\bm{k},\mu)\langle \partial_{k_{\nu}} u_{\alpha\bm{k}}|u_{\beta\bm{k}}\rangle \langle u_{\beta\bm{k}}|\partial_{k_{\rho}}u_{n\bm{k}}\rangle,\\
    \label{N3_plus_boundary}
\end{align}
with
\begin{equation}
    \mathcal{F}_{xy}^{\alpha}(\bm{k})=i(\langle \partial_{k_x}u_{\alpha\bm{k}}|\partial_{k_y}u_{\alpha\bm{k}}\rangle-\langle \partial_{k_y}u_{\alpha\bm{k}}|\partial_{k_x}u_{\alpha\bm{k}}\rangle) 
\end{equation}
the Berry curvature of the $\alpha$-th Bloch band and
\begin{align}
\notag
    \mathcal{B}_{\alpha\beta}(\bm{k},\mu) =& \int\!\!dz e^{z0^{+}}\!\frac{\partial}{\partial z}(g_{\alpha\bm{k}}g_{\beta\bm{k}}^{-1})\\
    =& -2 i\textrm{Im}[g_{\alpha\bm{k}}(\mu+i0^{+})g_{\beta\bm{k}}^{-1}(\mu+i0^{+})]
    \label{Bound_cont}
\end{align}
an on-shell (boundary) term that only depends on the properties of the Green's function at the chemical potential. If there are no poles nor zeros at $\mu$, this term vanishes identically and Eq.~\eqref{N3_plus_boundary} reduces to Eq.~\eqref{N3_ansatz} in the main text.

As a sanity check, one can evaluate Eq.~\eqref{N3_plus_boundary} in the non-interacting limit, where $\hat{G}_{\bm{k}}(z)=(z-\hat{H}_{\bm{k}})^{-1}$, with $\hat{H}_{\bm{k}}$ the single-particle Bloch Hamiltonian. In this case, $g_{\alpha\bm{k}}(z)=1/(z-\varepsilon_{\alpha\bm{k}})$, with $\varepsilon_{\alpha\bm{k}}$ the energy of the Bloch band corresponding to the eigenstate $|u_{\alpha\bm{k}}\rangle$. The boundary contribution in Eq.~\eqref{Bound_cont} then evaluates to
\begin{equation}
    \mathcal{B}_{\alpha\beta}(\bm{k},\mu) = -2\pi i(\varepsilon_{\bm{k}\beta}-\varepsilon_{\bm{k}\alpha})\delta(\varepsilon_{\bm{k}\alpha}-\mu),
\end{equation}
and Eq.~\eqref{N3_plus_boundary} reduces to
\begin{align}
\notag
    N_3[\mathrm{G}] =& \frac{1}{2\pi}\sum_{\alpha} \int d^2k \mathcal{F}_{xy}^{\alpha}(\bm{k})\Theta(\mu-\varepsilon_{\alpha\bm{k}})\\
    +&\frac{\Phi_0}{(2\pi)^2}\sum_{\alpha} \int d^2k m_z^{\alpha}(\bm{k})\delta(\varepsilon_{\alpha\bm{k}}-\mu),
    \label{non-int}
\end{align}
where
\begin{equation}
    m_z^{\alpha}(\bm{k}) = \frac{2\pi}{\Phi_0}\textrm{Im}\left[\langle \partial_{k_x} u_{\alpha\bm{k}}|\hat{H}_{\bm{k}}-\varepsilon_{\alpha\bm{k}}|\partial_{k_y}u_{\alpha\bm{k}}\rangle\right]
\end{equation}
is the intrinsic orbital magnetic moment of the $\alpha$-th Bloch eigenstate. This reproduces the known St\v{r}eda response of non-interacting Bloch electrons~\cite{PeraltaGavensky2025,Fregoso2026}, for which $N_3[\mathrm{G}]=\Phi_0 \partial n/\partial B$ at fixed chemical potential and zero temperature, explicitly given by Eq.~\eqref{non-int}.


\end{document}